\documentclass[11pt,a4paper]{article}

\newcommand{\rd}{{\rm d}}
\usepackage{amsmath,amsthm,amssymb}
\usepackage{graphicx}
\usepackage{epsfig,cite}
\usepackage[dvips]{color}
\usepackage{multicol}
\makeatletter
\@addtoreset{equation}{section}
\makeatother

\begin{document}
\titlepage

\begin{flushright}
QMUL-PH-07-06\\
CERN-PH-TH/2007-043
\end{flushright}
\vspace{1cm}
\begin{center}
\Large\textbf{IR Inflation from Multiple Branes}\\
\vspace{2cm} 
\small\textbf{Steven Thomas}$^{a,}$\footnote{s.thomas@qmul.ac.uk}
\small\textbf{and John Ward}$^{a, b,}$\footnote{j.ward@qmul.ac.uk}\\
\end{center}
\begin{center}
$^a$\emph{Center for Research in String Theory, Department of Physics,\\
Queen Mary, University of London \\
Mile End Road, London, E1 4NS, U. K}\\
\vspace{0.5cm}
$^b$\emph{Theory Division, CERN \\
1211 Geneva, Switzerland}
\end{center}
\vspace{0.1cm}
\begin{abstract}
In this paper we examine the IR inflation scenario using the DBI action,
where we have $N$ multiple branes located near the tip of a warped geometry.
At large $N$ the solutions are similar in form to the more traditional
single brane models, however we find that it is difficult to simultaneously satisfy
the WMAP bounds on the scalar amplitude and the scalar spectral index.
We go on to examine two new solutions where $N=2$ and $N=3$ respectively, which both
have highly non-linear actions. The sound speed in both cases is dramatically different
from previous works, and for the $N=3$ case it can actually be zero. We show that inflation is
possible in both frameworks, and find that the scalar spectral index is bounded from above by unity.
The level of non-gaussian fluctuations are smaller in the $N=2$ case compared to the single brane
models, whilst those in the $N=3$ case are much larger.
\end{abstract}
\newpage
\tableofcontents
\newpage

\section{Introduction}
Although inflation is generally accepted as
a model for the early universe, it is still a paradigm lacking an explicit
derivation from a more fundamental theory such as string theory.
Conversely we expect that inflationary cosmology provides an important
testing ground for the inflationary models from string theory thanks to
the abundance of precision data~\cite{wmap}.
In recent years we have witnessed a massive increase in the number of
string-cosmology papers (see \cite{lectures} and the references therein).

One of the interesting and purely stringy models is the Dirac-Born-Infeld (DBI)
inflation~\cite{tong} in type IIB string theory. 
This relies on the existence of a mobile D3-brane
moving along the throat geometry of a warped flux compactification.
The inflaton arises from an open string mode representing the
relative distance of the mobile brane with respect to some fixed background.
The non-linear form of the DBI action in this warped background
was shown to lead to substantial inflation despite the fact that the brane
could be moving at near relativistic velocity. It was further shown
that there are generically two models of this kind of inflation which were
dubbed the UV and IR models~\cite{irinflation}.\footnote{See also \cite{wilsoninf}
for an interesting alternative.}
In the UV model, the brane moves towards an anti-brane located near
the bottom of a warped throat. The anti-brane would be localised near
the bottom of the throat since it screens the RR flux present in the
compactification, and this corresponds to a form of
large field inflation. The IR model, in contrast, arises after brane flux
annihilation where mobile branes travelling down the throat annihilate
quantum mechanically with the trapped flux creating new branes in the process \cite{giantinflaton}.
These branes would then feel the attractive force of branes/fluxes in other
throats and so would generically be attracted towards the gluing region.
This corresponds to a small field inflation model. In fact the most general
case is an interpolating one, where one considers the brane starting in the UV,
travelling down the throat, annihilating with the flux and then
residual brane returning back up the throat.
It was shown that the level of non-gaussianities predicted
in this model are proportional to the square of the relativistic velocity factor.
Since this is usually greater than unity in these kind of models, we can
falsify models of this type by finding only small non-gaussianities
in future experiments.

It seems unlikely that only a single residual brane emerges from the flux annihilation
process, since this would require extreme fine tuning of the background fluxes.
More generally one would assume that there will be $N$ branes left after the flux annihilation,
and they all feel the same attractive force driving them back
up the throat. Indeed this was considered in \cite{irinflation}.
However here it is supposed that the branes are all separated by
distances greater than the string length, which just gives rise to a modified version
of assisted inflation \cite{assisted}. Since the
branes are all created in the same annihilation process, it is natural
that they are located at various distances, with some being coincident.
For simplicity we consider that all the branes are basically coincident
at the bottom of the throat. We can then describe their dynamics by
the non-Abelian DBI action, which is given by the Myers
action~\cite{myers}.\footnote{There is also a proposal by
Tseytlin~\cite{tseytlin} for the non-Abelian action, but
we use the Myers formalism in this paper.}
This action is complicated but simplifies when one takes the large $N$ limit. Although this appears to introduce
large dimensionless parameters into the theory, it can still be useful for inflationary model building \cite{nflation, giantinflaton}.

Another motivation for considering $N$ brane configuration is in relation
to reheating \cite{reheating}. In the usual single brane models, it is assumed that
the Standard Model (SM) lives in a separate throat from the inflationary throat.
It is supposed that once inflation ends, reheating occurs in the SM throat
through graviton production which tunnels through the internal space.
However in the absence of brane annihilation at the end of inflation,
there is still a residual massless $U(1)$ degree of freedom
in the inflationary throat. Then the reheating energy can be transferred
to the $U(1)$ gauge boson rather than to the SM. By contrast,
with multiple branes all the reheating energy couple directly to
the SM degrees of freedom. This is an interesting possibility which deserves
further investigation, but will be left for future study.

Since the DBI model of inflation is entirely stringy in origin, it is in our interest to
scan as much of the possible parameter space of solutions as possible. This will tell us more about
the robustness and predictiveness of the model, and also whether this kind of inflation is generic.
There has been recent work in this direction, however we hope to probe more of the parameter
space by considering multiple branes as they are typically more common in string theory.

In section 2 we introduce the non-Abelian action and the relevant backgrounds.
In section 3 we consider the simplest approximation of the the large $N$ limit
for the action. This shares many of the features of single brane models
while retaining its inherently non-Abelian structure. We investigate
the inflationary parameter space and determine the level of the curvature and
gravitational perturbations.
In section 3, we discuss the case of finite $N$. To highlight the differences
between multi and single brane inflation, we examine the cases of $N=2$ and $N=3$
in some detail. We calculate the speed of sound in both cases, and use this
to constrain the parameter space. We also calculate the relevant inflationary
parameters and the perturbation amplitudes and non-gaussianities of the model.
We finish with some conclusions and suggestions for future work.

The results of this paper can be summarized as follows. 

In the context of the non-Abelian world volume theory of $N$ $D3$-branes, 
we find, in the large $N$ limit, similar solutions to those of the single brane scenarios previously considered,
with the additional constraint that the background charges must be larger than the number of branes.
Our model differs significantly in some respects from a single probe brane,
despite sharing many similarities. The main
difference is that now the world-volume theory is playing the role of
the universe, rather than just the inflaton sector in the standard
IR approach to inflation. In fact this makes the model a hybrid between that
of DBI inflation \cite{tong} and Mirage Cosmology \cite{mirage}.

With certain assumptions, we have found that inflation is reasonably generic, i.e. we can generate at least
$60$ e-folds with relative ease due to large $N$ suppression.
We investigate in detail the cases of Ads and mass-gap backgrounds and find that the level of non-gaussianities are the same 
and that spectral indices are positive running and bounded from above by unity.
This large $N$ model may, however, be ruled out by experiment as that the large factors of $N$ tend to enhance the 
size of scalar perturbations. 

The case of finite $N$ is particularly interesting and striking.
For $N=2$ we find that the non-gaussianities are suppressed a little relative to the single brane case.
The extra stringy degrees of freedom present in the theory serve to induce a so-called
'fuzzy' potential term which makes the form of the action for finite $N$
significantly different from that of a single brane. 
As a consequence we find that in the relativistic approximation, the speed of sound
becomes a rather complicated function at finite $N$ and we are restricted to certain
regions of the moduli space of solutions.
For the $N=3$ case we even find that the speed of sound actually has zeros in its function.
The physical explanation for this is not immediately obvious to us, but is clearly a signature of the 
underlying complexity of the non-Abelian world volume theory.

\section{The non-Abelian action}
In this section, we introduce the non-Abelian action~\cite{myers},
which provides an effective description of coincident D-branes.
We first discuss the string theory background relevant for our model.

Let us assume that the type IIB spacetime metric factorises as follows
\begin{equation}\label{eq:metric}
ds^2 = h^2 ds_{4}^2 + h^{-2}(d\rho^2 + ds_{X_5}^2),
\end{equation}
where the four-dimensional metric is taken to be the usual
Friedmann-Robertson-Walker (FRW) form characterised by the scale factor $a(t)$,
a throat region is over some five-dimensional manifold $X_5$, and $h$ is a warp
factor.
We take the so-called IR inflation, where the branes are initially
localised at the IR tip of the throat where the warp factor is small.
These IR localised branes arise naturally in the context of brane-flux
annihilation.

Now consider a background where we have $M$ units of D3-brane
flux threading some three-cycle of the internal space.
If we insert $N'$ $\bar{D}3$-branes into this background
such that they fill the large $3+1$ dimensional spacetime, they
will feel an attractive force from the flux and roll down the throat to
annihilate quantum mechanically. Then there will be precisely $N=M-N'$
D3-branes created after this annihilation process. This provides the
initial conditions for our model.
In realistic string compactifications, there are many throats,
and their respective orientifold images. Therefore it is natural
that these remnant branes will feel attractive force from anti-branes
in other throats. Thus the remnant branes will travel along the
throat towards the internal space.

This model has been investigated recently for a single brane, $N=1$.
However this means that we need fine tuning to ensure that there is only
a single brane remaining after flux annihilation.
In a more generic situation, several branes will remain after this process.
For simplicity, we consider the case in which the branes are at distances
smaller than the string scale with $U(N)$ symmetry. Then we should use
the non-Abelian DBI effective action which describes coincident branes.

The bosonic components of the Myers action~\cite{myers}:
\begin{equation}\label{eq:myersaction}
S = -T_3 \int d^4 \xi {\rm STr} \left(\sqrt{-{\rm det}(\hat{E}_{ab}
+\hat{E}_{ai}(Q^{-1}-\delta)^{ij}\hat{E}_{jb}+\lambda F_{ab})}
\sqrt{\rm{det}Q^{i}_j} \right).
\end{equation}
The fields $\hat{E}_{ab}$ are the non-Abelian pullback of the linear
combination of closed string fields $E_{ab} = G_{ab} + B_{ab}$, while
the matrix $Q$ is determined as
\begin{equation}
Q^i_j = \delta^i_j + i\lambda [\phi^i, \phi^k] E_{kj}\,,
\end{equation}
with the $\phi^i$ being scalar fields on the world-volume of the D-branes
corresponding to the transverse fluctuations and $\lambda = 2\pi l_s^2$ with $l_s$ the 
fundamental string length. Implicitly the scalars are now
$N \times N$ matrices transforming in the irreducible representation of
their respective gauge group. This irreducibility condition essentially means that
they are in their lowest energy state.
The symmetrised trace prescription means that we must trace over the
symmetric average of all the fields in the action.
Building upon recent work \cite{warpeddeformedconifold, blowup} we assume that the bulk $B$ fields is zero near
the tip of the throat, and consider the case where the transverse
coordinates define a fuzzy $S^2$ embedded in a three cycle in the $X_5$ manifold.

A fuzzy sphere is defined in a similar way to an ordinary sphere, except
that the defining relation for the $d$-dimensional sphere 
\begin{equation}
\sum_{i=1}^{d+1} (x^i)^2 = R^2, \hspace{2cm}
x^i \in \mathbb{R}^{d+1},
\end{equation}
is subject to the replacement of $x^i$ by $N$-dimensional matrices $\hat{x}^i$.
Any function expanded in terms of spherical harmonics
on the sphere is subject to a cut-off due to the finite size of the matrices.
Mathematically this means that we are truncating the algebra of functions on
the sphere.
Thus a fuzzy sphere is essentially a sphere where we can identify only $N$ points.
For the two-sphere, it is easy to see that in the large $N$ limit,
the fuzzy $S^2$ coincides with the classical two-sphere.
Note that this is not generally the case for higher dimensional fuzzy spheres.

Now that we have restricted ourselves to the $SO(3) \sim SU(2)$ algebra,
we expect the scalars to be proportional to its generators.
As argued in \cite{myers}, it is preferable for the generators to be
in the irreducible representation as this corresponds to the lowest energy
configuration.
The matrix ansatz that we take for our scalar fields is thus $\phi^i = R \alpha^i$
($i=1, 2, 3$), where $R$ is a variable of canonical mass dimension,
and the $\alpha^i$ are the $N$-dimensional generators of the $SU(2)$ algebra.

In flat space, we can find solutions where the radius of the fuzzy sphere
grows without bound. However for most compactifications there will be a maximum
bound on the size of the fuzzy sphere, as illustrated by the famous example
of the AdS$_5 \times S^5$ metric where the fuzzy sphere radius is bounded by
the radius of the five-sphere. In the warped backgrounds where the length
of the throat provides the cutoff scale, the throat is smoothly glued onto
the five-dimensional internal space. We call this the UV cut-off, analogous
to the UV brane in the Randall-Sundrum models \cite{randallsundrum}.

Let us insert the metric (\ref{eq:metric}) into the Myers action
(\ref{eq:myersaction}).
After expanding all the determinants, we find that the effective action
for coincident D3-branes in this background\footnote{This is based upon earlier work \cite{fuzzycurved}.} becomes 
\begin{equation}\label{eq:action}
S=-T_3 \int d^4 \xi {\rm STr} \left(h^4 a^3 \sqrt{1-h^{-4}\lambda^2
\alpha^i \alpha^i\dot{R}^2}
\sqrt{1+4\lambda^2 \alpha^i \alpha^i h^{-4}R^4}- a^3 h^4{\bf 1_N}
+ a^3 V(R){\bf 1_N} \right),
\end{equation}
where the second term arises as the leading order contribution from the
Chern-Simons coupling of the bulk RR fields, while the final term is
a flux induced potential which comes from fluxes present in
the background. Both terms are singlets under the trace, which
is why they appear multiplied by the $N \times N$ identity matrix.
The potential generated will depend on the topology of the
internal space, and also the quantisation constraints of the fluxes.
Here we have absorbed a factor of $T_3^{-1}$ into the potential to
make it dimensionless, and set the dilaton to be unity. This agrees with
the supergravity background generated by $D3$-branes, and also the tip solution
of the Klebanov-Strassler throat \cite{klebanovstrassler, warpeddeformedconifold}.

We see that the non-linear form of the DBI action leads to an infinite
series expansion in powers of the $SU(2)$ generators. The action is thus
complicated by the symmetrised trace over the generators.
The action can be simplified by keeping only the leading
term in each of the square roots, which is the large $N$ limit.
In this approximation, all the terms in the action are proportional
to the identity matrix and
${\rm Tr}( \alpha^i \alpha^i) = \hat{C} {\rm Tr} (\mathbf{1})$ where $\hat{C}$
is the quadratic Casimir of the $N$-dimensional representation of $SU(2)$.

We now minimally couple the DBI action to the standard Einstein-Hilbert action.
Fortunately the Einstein part of the action arises
naturally in the process of compactification, so we can consistently include it in our model.
It is convenient to introduce the notation
\begin{equation}
W(R,\hat{C}) = \sqrt{1+4\lambda^2 h^{-4}\alpha^i \alpha^i R^4},
\end{equation}
which is essentially an additional potential induced by the fuzzy sphere geometry,
which we will refer to as the fuzzy potential.
For a single probe brane, this contribution is always unity, since the
corresponding matrix representation is commutative.
We can now determine the non-zero components of the
energy-momentum tensor on the world-volume. It gives
the energy and pressure densities as
\begin{eqnarray}\label{eq:general}
E &=& T_3 {\rm STr} \left(\frac{W(R, \hat{C})h^4}{\sqrt{1-h^{-4}\lambda^2
\alpha^i \alpha^i \dot{R}^2}} - h^4 +V \right), \\
P &=& - T_3 {\rm STr} \left(h^4 W(R,\hat{C})\sqrt{1-h^{-4}\lambda^2 \alpha^i
\alpha^i \dot{R}^2} -h^4 +V \right).
\end{eqnarray}
These form the basis for all our analysis in the subsequent
sections and we will return to them in due course.

One of the reasons that we can have DBI inflation in this background is that
the speed of sound is usually very small compared to unity.
In the standard canonically normalised slow roll models, this factor is
always unity since the scalar field moves slowly.
The speed for the non-Abelian case is
\begin{equation}\label{eq:speedofsound}
C_s^2 = \frac{\partial P/\partial \dot{R}}{\partial E/\partial \dot{R}},
\end{equation}
which is valid provided that the entropy corrections are negligible.
\section{The large $N$ limit}
In this section, we focus on the large $N$ limit. We show that the solution
shares many of the features with the single brane models.

Let us restrict ourselves to the large $N$ limit when we ignore
the backreaction\footnote{This amounts to a constraint on the
energy of the D-branes which could lead to
an unphysical solution upon compactification. However it
can be interpreted as a metric constraint on the bulk fluxes.}.
The large $N$ limit has proven to be useful for inflation in many other
contexts~\cite{giantinflaton, nflation}.
Now we can approximate the symmetrised trace by a trace once we neglect
the contributions of terms of the order $1/N^2$ in the action.
The resulting action simplifies to
\begin{equation}\label{eq:largeNaction}
S = -T_3 \int d^4 \xi N a^3 \left(h^4 \sqrt{1-h^{-4}\lambda^2 \hat{C} \dot{R}^2}
\sqrt{1+4\lambda^2 \hat{C} h^{-4} R^4} -h^4 + V(R) \right)
\end{equation}
where $T_3$ is the tension of the $D3$-branes given by
\begin{equation}
T_3 = \frac{M_s^4}{8\pi^3 g_s}
\end{equation}
with $M_s$ being the mass scale for the open strings and $g_s$ being the asymptotic string coupling which we take to be small to
allow for perturbatively defined strings. In our paper we will assume that the coupling is set to $g_s \sim 10^{-2}$ in order to
make order of magnitude approximations.
The back reaction effect will be small provided that we ensure $MK>>N$ is also satisfied,
in addition to the large $N$ assumption.
It is instructive to make a redefinition of the scalar field.
Firstly we switch to 'physical' coordinates using the relation
$R^2 =r^2/(\lambda^2 \hat{C})$
which parameterises the physical radius of the fuzzy sphere.
This defines the \emph{square} of the physical radius.
Let us also define a scalar field $\phi = r \sqrt{T_3}$ with canonical mass
dimensions in order to make contact with the rest of the literature.

We have the standard relationship between the four-dimensional Planck
scale and the ten dimensional one through the volume of the warped space $V_6$:
\begin{equation}
M_p^2 = \frac{V_6}{\kappa_{10}^2}.
\end{equation}
Interestingly it was shown in \cite{braneconstraints} that with minimal
assumptions about the volume of the throat, one can find the following
bound on the maximal field variation:
\begin{equation}
\Delta \phi < \frac{2 M_p}{\sqrt{MK}},
\end{equation}
where $MK$ is the contribution from the background fluxes.
In our case we demand that $MK >> N$ to neglect the
back reaction upon the geometry, which restricts our field to move over
very small distances in Planckian units.

We find from (\ref{eq:myersaction}) the equation of motion of the $\phi$ field:
\begin{eqnarray}
0 &=& W(\phi, \hat{C}) \gamma^3 \ddot{\phi} + 3H\dot{\phi}W(\phi, \hat{C}) \gamma
 + \frac{8 \gamma \phi^3}{T_3 \lambda^2 \hat{C} W(\phi, \hat{C})}
 \left(1-\frac{\phi h'}{h} \right) \nonumber\\
&+& 2T_3h' W(\phi, \hat{C})\gamma \left(2h^3 - \frac{\gamma^2
\dot{\phi}^2}{h T_3} \right) - 4 T_3 h^3 h' + T_3 V'\,,
\end{eqnarray}
where primes are derivatives with respect to $\phi$ and we
have also introduced
\begin{equation}\label{eq:gamma}
\gamma = \frac{1}{\sqrt{1-h^{-4}T_3^{-1}\dot{\phi}^2}},
\end{equation}
for the analogue of the relativistic factor for the DBI action. This implies
that the velocity of the brane is bounded as
\begin{equation}
\dot{\phi}^2 < h^4 T_3.
\end{equation}
These last two equations are exactly the same as in the case of a single brane \cite{tong}.
Recall that when we take the large $N$ limit of the fuzzy $S^2$,
we recover the classical $S^2$ with $N$ units of charge.

Using the general expression~(\ref{eq:speedofsound}), we calculate the speed
of sound $C_s^2=1/\gamma^2$, in agreement with that of single brane inflation.
As in the single brane model, we now assume that the scalar field is monotonic,
at least for the early times. This assumption allows us to switch to the 
Hamilton-Jacobi formalism, whereby we use the field itself
as the dynamical parameter rather than time variable.
We differentiate the Friedmann equation $H^2= E/3M_P^2$ with respect
to time, dropping terms proportional to $\ddot{\phi}$, and use
the continuity equation, $\dot{E} = -3H(P+E)$ to get
\begin{eqnarray}\label{eq:HJ_largeN1}
\dot{\phi} = -\frac{2M_p^2 H'}{N\gamma W(\phi)}\,,
\end{eqnarray}
where the fuzzy potential $W$ is now an explicit function of $\phi$, and $H'$
is the derivative of the Hubble parameter with respect to the inflaton.
Substituting this $\dot \phi$ into (\ref{eq:gamma}), we obtain
\begin{eqnarray}\label{eq:HJ_largeN2}
\gamma(\phi) = \sqrt{1+\frac{4M_p^4 H'^2}{N^2 W^2(\phi) h^4 T_3}} \,.
\end{eqnarray}
We can use~(\ref{eq:HJ_largeN2}) to write the velocity
of the inflaton without reference to the relativistic factor $\gamma$:
\begin{equation}
\dot{\phi} = \frac{-2M_p^2 H'}{\sqrt{N^2 W^2 + 4M_p^4 H'^2 h^{-4} T_3^{-1}}},
\end{equation}
which allows us to write the corresponding Friedmann equation solely as
a function of the inflaton:
\begin{eqnarray}\label{eq:master}
H^2 &=& \frac{N T_3}{3M_p^2}\left(W(\phi)h^4(\phi) \gamma(\phi)
+ V(\phi) -h^4(\phi) \right) \\
&=& \frac{NT_3}{3M_p^2}\left(V(\phi) + h^4(\phi) \left\lbrace \frac{1}{N}
\sqrt{N^2 W^2 + \frac{4M_p^4 H'^2}{h^4 T_3}}-1 \right\rbrace \right), \nonumber
\end{eqnarray}
which is reminiscent of the equation found in \cite{tong} for
the Abelian DBI inflation model. The main difference here is
the presence of factors of $N$ and the fuzzy sphere induced potential
$W(\phi)$, the latter being an inherently non-Abelian feature.
In this manner we have written the variables purely as functions
of the scalar field.

We may be concerned that the DBI models of inflation do not exhibit standard
attractor solutions for inflation, since we expect relativistic motion.
To check this, let us suppose that $H_0(\phi)$ is a solution of~(\ref{eq:master}), which can be either inflationary or non-inflationary.
We add to this a linear homogeneous perturbation $\delta H(\phi)$.
The attractor condition will be satisfied if it
becomes small as $\phi$ increases.
Upon substituting $H= H_0 + \delta H$ into (\ref{eq:master})
and linearizing the resultant expression, we find that the perturbation obeys
\begin{eqnarray}
H'_0 \delta H' = \frac{3NW\gamma}{2M_p^2}H_0 \delta H,
\end{eqnarray}
which has the general solution
\begin{eqnarray}
\delta H (\phi) = \delta H (\phi_i) \exp \left[\frac{3N}{2M_p^2}
\int_{\phi_i}^{\phi}d\phi
W(\phi)\gamma(\phi)\frac{H_0(\phi)}{H'_0(\phi)} \right],
\end{eqnarray}
where $\delta H(\phi_i)$ is the value at some initial point
$\phi_i$, and $\gamma = \gamma(H_0)$. Because $H'_0$ and $d\phi$ have
opposite signs, the integrand within the exponential term is negative
definite, and all linear perturbations indeed die away.
This means that there is an attractor solution for this model regardless of the
initial conditions and the velocity of the brane.
This is also true for the single brane solutions~\cite{irinflation} with
$N=1=W(\phi)$.

We note that the equation of state in this model is drastically different from
the canonical field models. It is for the large N
\begin{equation}\label{eq:eos}
\omega = - \frac{Wh^4 \gamma^{-1} + V(\phi) - h^4}{Wh^4 \gamma + V(\phi) -h^4}.
\end{equation}
If the potential dominates all the other terms, we recover the usual
de-Sitter solution with $\omega \sim -1$.
However the DBI admits more interesting solutions due to its non-linear nature.
For example, if we consider ultra-relativistic motion where $\gamma >>1$,
and demand that the $h^4$ terms are suppressed, we obtain
\begin{equation}
\omega \sim \frac{-V(\phi)}{Wh^4 \gamma + V(\phi)},
\end{equation}
which can be very small depending on the scale of the fuzzy potential,
and may give rise to a matter phase in the asymptotic velocity limit.
This shows that we have a larger parameter space of solutions for $\omega$
than in the standard inflationary scenarios.

Our analysis thus far has been general. To make more detailed investigation
of the inflationary signature of this model, we must determine the background
potential.
Let us consider nonzero fluxes inducing the warped throat solution.
The coincident branes localised at the bottom of the throat will feel
a potential generated by branes/fluxes in other throats and will move
towards them. We then expect a tachyonic potential of the form
\begin{equation}
V(\phi) \sim V_0 - \frac{V_2 \phi^2}{2} + \ldots,
\end{equation}
with higher order even powers of $\phi$ because of the $\mathbb{Z}_2$
symmetry of the throat.
The various constants will be determined by the choice of
compactification, and also non-perturbative effects.
The IR DBI inflation is thus a special case of small field inflation.
The constant $V_0$ is the scale set by the fluxes, and
need be large to be able to neglect back-reactive effects in our model.
This is significantly different from the UV inflationary model.
In this paper we are mainly interested in the IR solution,
and we consider that most of the dynamics will take place in a
region dominated by the potential energy.

\subsection{Inflationary observables and constraints}

In this subsection, we focus on inflationary solutions in two specific
backgrounds. We show that inflation can be achieved in this model
by allowing an appropriate tuning of the parameters in the theory.

We wish to consider inflation near the tip of a warped throat where $\phi$
is small. However we are also interested in solutions where
the background is no longer approximately AdS but is constant.
This is motivated by the work of \cite{braneconstraints}
and also the fact that a finite warp factor appears to be a generic
feature of the IR end of warped throats. This is a phenomenologically motivated
solution, because once we compactify our theory on a compact space, we expect
the fluxes to back-react on the bulk geometry forming a throat.
We expect that these throats will be of the Klebanov-Strassler (KS)
variety, which have a finite cut-off at the origin. This cut-off is
generically exponentially small due to its dependence on the three-form
fluxes.

In order to mimic a constant warping in our non-compact theory
we choose to put in 'by hand' a constant warp factor parameterised by
a mass term $\mu$, where we expect that $\mu$ is small at a
scale set by the bulk fluxes:
\begin{equation}
h(\phi) = \frac{\sqrt{\phi^2+\mu^2}}{L},
\end{equation}
where we have used $L$ to denote the background charge.
It should be noted that this $L$ will be
different from that in the AdS-like backgrounds.
When $\phi$ goes to zero, the warp factor remains finite.
Strictly speaking away from the origin there may be a different $\phi$ dependence,
but we use this form of the warp factor in the following sections.

The solutions we consider are\\
(i) the AdS type cases (where we set $\mu$ to 0), or \\
(ii) the mass gap cases (where we assume $\phi \sim 0$).

Let us assume that inflation occurs very close to the tip of the warped throat,
in which case we expect the energy density to be dominated by the
constant piece of the background potential. The other terms are suppressed
by the square of the warp factor and will be small in this limit.
It may appear that the warp factor can be easily vanishingly small,
but more care is required in cases where $h$ reduces to a constant, since
in those backgrounds (as we shall see) the other parameters can \emph{both}
be large. Assuming $V_0 >> h^4(W(\phi)\gamma(\phi)-1)$ is satisfied,
the Friedmann equation~(\ref{eq:master}) can be approximated as
\begin{equation}\label{eq:largeNhubble}
H^2 \sim \frac{NT_3 V(\phi)}{3M_p^2}.
\end{equation}
We then find that our solutions for the inflaton velocity and
the gamma factor reduce to
\begin{eqnarray}
\dot{\phi} &\sim& -\frac{M_p V'}{\gamma W}
\sqrt{\frac{T_3}{3NV}} \,, \nonumber \\
\gamma(\phi) &\sim& \sqrt{1+\frac{M_p^2}{3N} \left(\frac{V'^2}{h^4
W^2 V} \right)} \,,
\end{eqnarray} where we have expressed everything
in terms of derivatives of the potential. Clearly these
expressions are background dependent as they depend on the warp factor $h$.
\subsection{Inflation in AdS type backgrounds}

Let us consider
the locally AdS type backgrounds, such as those studied in the
original IR inflation scenario~\cite{irinflation}.
The harmonic function can be approximated by $h \sim \phi/L$ in the near
horizon region where $L$ is essentially the charge of the background
geometry given by\footnote{Note we have
rescaled this quantity to ensure it has the correct dimensions.
This corresponds to the radius of curvature for the AdS space
scaled by the square of the brane tension} $L^4 = g_s M K T_3^2
\lambda^2 \pi^2/ \rm{Vol(X_5)}$, with $M$ and $K$ the
corresponding quanta of flux and $\rm{Vol(X_5)}$ corresponding to
the dimensionless volume of the compact space. We expect
to find that $\rm{Vol(X_5)}=a \pi^3$, where $a$ is a topological
parameter. For example we know that $a=1$ for the five-sphere and
$a=16/27$ for the manifold $T^{1,1}$ \cite{braneconstraints}. 
In this situation the fuzzy sphere induced potential becomes constant
which greatly simplifies the analysis. In fact the second term
inside the square root is proportional to the ratio of the
background fluxes and the number of coincident branes, and thus it
is not clear a priori whether this term will be small or large. Under
our assumption of no back reaction, we require the flux
term to dominate over $N$, and so we may expect that $W >> 1$
which translates into the flux condition
\begin{equation}
\frac{L^2}{M_s^2} >> \frac{N}{8 \pi^2 g_s} \hspace{0.5cm}\to \hspace{0.5cm} \sqrt{MKg_s} >> N
\end{equation}
in the large $N$ limit. For small values of the string coupling constant,
we see that this requires the fluxes to be very large.
This is to expected from out heuristic arguments regarding the backreaction.
In the converse limit where $W \sim 1$ we see that the constraint becomes $\sqrt{g_s MK} << N$ - which although provides a bound on $N$ is much
harder to satisfy within the remit of our approximation. 

This requirement also affects the definition of $\gamma$:
\begin{equation}
\gamma_{\rm{AdS}}(\phi) \to \sqrt{1+ \frac{M_p^2 V_2^2 L^4}{3NW^2V_0 \phi^2}
\left(1+\frac{V_2 \phi^2}{V_0} + \ldots \right)}.
\end{equation}
As we are interested in DBI inflation, we should take $\gamma(\phi) >>1$ for
the speed of sound to be substantially reduced, and so the
right hand side will dominate the previous expression.
Furthermore since we are assuming that the background potential should dominate the
energy density at this stage of the evolution, we should also assume that
$V_0 >> V_2$, which implies that the new constant piece of $\gamma$
will be subdominant.

In this regime, we can approximately solve the inflaton equation of motion.
In fact there is a cancellation between terms in the equation which implies that
$\dot{\phi} \sim \mathcal{O}(\phi^2) + \mathcal{O}(\phi^6)$ and so
for consistency we must drop all terms higher than quadratic in the fields.
The resultant expression for the field is actually of the same functional
form as the single brane case. Actually this is expected since we can view
the fuzzy sphere as a classical sphere with $N$ units of flux in the large $N$
limit:
\begin{equation}
\phi \sim \frac{\phi_0 L^2}{ L^2 - \phi_0 \sqrt{T_3}(t-t_0)},
\end{equation}
where the field is initially located at $\phi = \phi_0$ at $t=t_0$.
This expression on the denominator will generally be much smaller than
unity as we are assuming that $h_0 << 1$, which is the initial value
of the warp factor.

Let us now compute the inflationary parameters in this large $N$ limit.
The non-linear form of the DBI action prevents us from using the traditional
slow roll variables, and so we must establish new 'fast roll' variables.
In reality the name 'fast roll' is somewhat of a misnomer because despite moving relativistically,
the non-linear nature of the action allows for the brane to be held up on the potential for a significant
amount of time as in slow roll scenarios.
This has already been extensively discussed for the single brane models
but the non-Abelian action requires us to modify these expressions.
Suppose that the leading order term for the epsilon parameter expansion is given by
\begin{equation}
\frac{\ddot{a}}{a} = H^2(1-\varepsilon),
\end{equation}
which yields the usual slow roll constraint $\varepsilon = -\dot{H}/H^2$.
However now that we are working in the Hamilton-Jacobi formalism,
we need derivatives with respect to the inflaton. This leads to the
following modified expressions for the relevant slow roll parameter:
\begin{eqnarray}\label{eq:largeNslowroll}
\varepsilon &=& \frac{2M_p^2}{N\gamma W(\phi, \hat{C})} \left(\frac{H'}{H}
\right)^2. \vspace{0.5cm} 
\end{eqnarray}
This is clearly equivalent to the usual single brane slow roll conditions
where we would have $N=1, W(\phi, \hat{C})=1$, although
we cannot take the $N \to 1$ limit in our non-Abelian DBI description.
Note that this slow roll parameter is suppressed not only
by a factor of $1/\gamma$ as in the single brane inflation, but also by
an additional factor of $1/N$. Intuitively we may have expected this since the
coincident branes will tend to accelerate much more slowly than a single brane.
Hence we would expect inflation to last for a longer period of time.

Using our approximate solutions we find for the AdS type backgrounds
(and assuming $\gamma >> 1$)
\begin{equation}
\varepsilon \sim \frac{\sqrt{3} M_p V_2 \phi^3}{2L^2 V_0 \sqrt{NV_0} } + \ldots ,
\end{equation}
where we have neglected $\phi^5$ terms. We must ensure that $\varepsilon < 1$
for inflation to occur at all. We can also calculate
the number of e-foldings for this model using
\begin{equation}\label{eq:efoldings}
N_e = \int H dt.
\end{equation}
We obtain
\begin{equation}
N_e \sim \sqrt{\frac{NV_0}{3}}\frac{L^2}{M_p} \int d\phi
\frac{1}{\phi^2}\left(1-\frac{V_2 \phi^2}{4V_0} +
\mathcal{O}(\phi^4) \right),
\end{equation}
where the integration should be between $\phi_0$ and $\phi_f$, the latter
being determined though the fast roll parameter. Clearly for small $\phi $ the
first term will dominate the integral and so we drop the higher order terms.
The result is that the field value $N_e$ e-folds before the end of inflation
can be written as
\begin{equation}
\phi_0 \sim \frac{\phi_f L^2 \sqrt{N V_0}}{L^2\sqrt{NV_0}
+ \sqrt{3}\phi_f N_e M_p },
\end{equation}
which we can use to determine the perturbation spectrum.
For completeness, we write the fast roll parameter as a function of the number
of e-foldings:
\begin{equation}
\varepsilon \sim \left( 1 + N_e \left(\frac{6 M_p^2}{NL^4V_2} \right)^{1/3} \right)^{-3}.
\end{equation}

\subsection{Inflation in mass gap backgrounds}

The equation of motion for the inflaton in the AdS background is basically
the same as in the single brane case. Is this also true for the
mass gap solution? In this instance the fuzzy sphere potential now has
non-trivial field dependence which complicates the analysis. Without loss of
generality we will take the warp factor of the form $h \sim \mu/L$:
\begin{eqnarray}\label{eq:massgap}
W_{\rm mg}(\phi)&\sim& \sqrt{1+ \frac{4\phi^4 L^4}{\mu^4 \lambda^2 N^2 T_3^2}},
\nonumber \\
\gamma_{\rm mg}(\phi) &\sim& \sqrt{1+\frac{M_p^2 L^4 V_2^2 \phi^2}{3N \mu^4
{W_{\rm mg}}^2 V_0}}.
\end{eqnarray}

In this paper we are mainly interested in the relativistic limits of the theory, and so we require $\gamma$ to be large.
This immediately imposes a constraint on the fuzzy potential - which is now an explicit function of the inflaton $\phi$
\begin{equation}\label{eq:mggammaconstraint}
\frac{M_p^2 V_2^2 \phi^2}{3V_0 N h^4} >> W^2.
\end{equation}
The analytic form of $W$ tells us that it is bounded from below by unity, but has no upper bound. Of course in reality 
we expect $W(\phi)$ to be a monotonically increasing function, however for analytical simplicity we will consider the two
limits separately.

In the first instance let us assume that $W\sim 1$. From (\ref{eq:mggammaconstraint}) this implies that we have an upper bound
on the number of branes
\begin{equation}
N << \frac{M_p^2 V_2^2 \phi^2}{3 V_0 h^4}
\end{equation}
which can be satisfied by having a small enough warp factor, and suitable hierarchy between terms in the potential. Note we are implicity
assuming that $\phi$ is non-zero here - since a vanishing field corresponds to a fuzzy sphere of zero size due to the relationship between
the canonically normalised scalar and the fuzzy sphere radius.
However we must also impose the constraint coming from the
definition of the fuzzy potential which actually implies a lower bound on $N$ through the relation
\begin{equation}
N >> \frac{2 \phi^2}{h^2 \lambda T_3}
\end{equation}
so upon combining these constraints we see that both assumptions are valid provided that
\begin{equation}\label{mgtotalconstraint}
\frac{V_2^2}{V_0} >> \frac{h^2}{M_s^2 M_p^2}
\end{equation}

We now ask about the constraints arising from the converse limit, when both $\gamma$ and $W$ are large. For the fuzzy potential
to be large we must ensure that
\begin{equation}
N << \frac{7 \times 10^{-1} \phi^2}{h^2 M_s^2},
\end{equation}
however the relativistic limit also requires $N$ to be bounded from below
\begin{equation}
N >> \frac{7 \times 10^{-1} \phi^2 V_0}{V_2^2 M_s^2 M_s^2}.
\end{equation}
Combining both of these constraints also results in (\ref{mgtotalconstraint}).
It is straight-forward to note that due to the constant warp factor near the tip of the throat, the relativistic limit implies
that $\phi$ is a linear function of time. This was already established for the single brane case by Kecskemeti et al \cite{uvmodels},
and follows from the fact that our assumption requires us to drop the factor of unity in the definition of $\gamma$, which 
then uniquely fixes the form of $\dot{\phi}$ to be constant.
The overall scale is simply set by the warp factor and the brane tension, and is independent
or our parameterisation of the fuzzy potential.

Let us now consider how inflation occurs in this model. Our starting point will once again be (\ref{eq:largeNslowroll}).
Inserting the mass gap solution and also demanding relativistic motion is enough to ensure that $W$ drops out of the analysis,
thus the solution is independent of the fuzzy potential contribution. After integrating to find the number of e-folds we see that
the slow roll parameter can be written in the form ($N_e$ e-folds before the end of inflation)
\begin{equation}
\varepsilon \sim 1-\frac{3N_e V_2 M_p^2 h^4}{2 V_0 N}
\end{equation}
We have seen that for a certain range of parameters, we can obtain the required
number of e-foldings. The real signature of the model lies in the perturbation
spectrum and the scale of the spectral indices, which we now address.

\subsection{Cosmological Signatures.}

The derivation of the various perturbation spectra for this model is presented in the Appendix, and we refer
the reader there to see how the expressions arise. We will simply quote the important results in the following section.
The main equations we need to calculate the perturbation spectra are
(\ref{eq:curvature}) and (\ref{eq:tensor}) respectively. In terms of our
standard notation employed in the rest of the paper these translate into
the following conditions when we substitute for the velocity equation
in the Hamilton-Jacobi formalism, and re-insert the factors of the reduced
Planck mass
\begin{eqnarray}
\mathcal{A}_S^2 &\simeq& \frac{H^2}{8 \pi^2 M_p^2 \varepsilon
C_s}\,,\nonumber \\
\mathcal{A}_T^2 &\simeq& 8 \left(\frac{H}{2 \pi M_p} \right)^2\,.
\end{eqnarray}
The slow-rolling inflation generally predicts very low
non-gaussianity since in the leading order the quantum fluctuation
are generated by free fields in the dS background. However in the
DBI inflation, much larger non-gaussianity can be generated since
the causality constraint in the kinetic term introduces non-linear
interactions among different momentum modes of the scalar field.
Recently, it was shown that in the equilateral triangle limit, the
leading-order contribution to the non-linearity parameter is given
by~\cite{braneconstraints}
\begin{eqnarray}
\label{nongauss}
 f_{NL} = \frac{35}{108}\left(\frac{1}{C_s^2}-1\right) -
\frac{5}{81}\left(\frac{1}{C_s^2}-1-2\Lambda \right),
\end{eqnarray}
where
\begin{eqnarray}
\label{lambda}
 \Lambda \equiv \frac{X^2 p_{,XX}+\frac{2}{3}X^3
p_{,XXX}}{X p_{,X} + 2X^2 p_{,XX}} \,.
\end{eqnarray}
It can be shown that the second term on the right-hand side of
~(\ref{nongauss}) vanishes in the large N case, as it does in the single brane case. 
Therefore we have the usual non-gaussian parameter
\begin{eqnarray}
f_{NL} \approx 0.32\gamma^2.
\end{eqnarray}
This once again emphasizes the similarity between the $N=1$ and the $N>>1$ descriptions. Current measurements
indicate a rather weak bound on the level of non-gaussianities $f_{NL} \le 100$, however the upcoming
Planck mission aims to increase the sensitivity to probe down to regions where $f_{NL} \le 5$.

We will now discuss the expected level of perturbations in each of the two cases we have discussed so far.
\begin{itemize}
\item \textbf{Ads backgrounds.}
Using the formulae derived in the appendix we find that the scalar amplitude at leading order becomes
\begin{equation}
\mathcal{A}_S^2 \sim \frac{NV_0}{12 \pi^5 g_s} \frac{\gamma}{16 \varepsilon} \left(\frac{M_s}{M_p} \right)^4.
\end{equation}
Now using the WMAP normalisation for the non-gaussianities we must ensure that $\gamma \le 10\sqrt{3}$, which in turn
fixes $\varepsilon$ through the (weak) constraint that $r \le 0.5$. Therefore we see that at horizon crossing we must ensure that
\begin{equation}\label{eq:epsilonconstraint}
\varepsilon \le \frac{5 \sqrt{3}}{16} \hspace{0.5cm}\to \hspace{0.5cm} MKN V_2 \le \frac{10^6 M_p^2}{M_s^4}
\end{equation}
is satisfied where we assume that the string coupling is roughly $10^{-2}$.
This essentially means that the slow roll parameter must be less than one half. Using this constraint to fix the
scalar amplitude we find the following condition needs to be satisfied
\begin{equation}\label{eq:tensorconstraint}
N V_0 \left(\frac{M_s}{M_p} \right)^4 \le 10^{-8}.
\end{equation}
Clearly this is linear in $N V_0$, therefore requires that the string scale must be low in order for the WMAP normalisation to hold,
assuming that the constant part of the potential is sub-Planckian.
In addition we see that the scalar index can be expanded as a power series in $\varepsilon$, which at leading order
becomes
\begin{equation}
n_s \sim 1 - 4\varepsilon - 4\varepsilon^{1/3} \delta n_s + \ldots
\end{equation}
where we have defined the 'perturbation'
\begin{equation}
\delta n_s \sim \left(\frac{10^2 M_p}{M_s^2 \sqrt{NV_2 g_s MK}} \right)^{2/3}.
\end{equation}
Recall that the WMAP bound for this index is given by $n_s = 0.987^{+0.019}_{-0.037}$. 
Using the constraint in (\ref{eq:epsilonconstraint}) we find that the $\delta n _s$ term must be bounded from below by unity.
Moreover we see that by constraining $\varepsilon$ to be smaller - this in fact makes the $\delta n_s$ term larger.
Given this, one sees that the scalar index will always be large and negative in this instance and therefore incompatible with
observation.

We also see using the relation 
\begin{equation}
\frac{d n_X}{d \ln{k}} \sim \frac{d n_X}{d N_e} \hspace{2cm} X= {S,T}
\end{equation}
evaluated at horizon crossing, that both tensor and scalar indices are positive  running. 
\item \textbf{Mass Gap Backgrounds.}
The form of the Hubble parameter in this instance is the same as in the Ads case, therefore we expect similar arguments to
hold in this instance. Once again this favours a smaller string scale using (\ref{eq:tensorconstraint}).
Assuming similar constraints on $r$ and $\gamma$ we see that the only real distinction between the two
cases arises through the spectral indices, since we again have to ensure that (\ref{eq:epsilonconstraint}) is satisfied.
Explicit calculation of the scalar index in this case reveals that
\begin{equation}
n_s -1 \sim -4\varepsilon \left(1 - \frac{H}{4H'} \left\lbrace\frac{3 H''}{H'} - \frac{2 W'}{W} \right\rbrace \right).
\end{equation}
If we restrict ourselves to the case where $W\sim 1$ then this simplifies down to the following
\begin{equation}
n_s \sim 1 - 7\varepsilon - \frac{9M_p^2 V_2}{2 V_0 N \varepsilon}.
\end{equation}
The scalar index will clearly be sensitive to the magnitude of the second term, which we write as $-\delta n_s/\varepsilon$, and we see that
\begin{equation}
\delta n_s \ge \frac{1}{20 h^4} \left(1-\frac{5 \sqrt{3}}{16} \right)
\end{equation}
using the WMAP normalisation. Clearly for small values of the warp factor we will see that the $\delta n_s$ term will be large, which
implies that we cannot obtain the observed spectrum of scalar perturbations. The only way we can satisfy the experimental data in this
instance is to assume a large warp factor - however this is contrary to all our assumptions thus far. So we must conclude that
this particular model does not agree with the data.

Conversely in the limit where we take $W>>1$ we find a cancellation between dangerous $1/\phi$ terms which gives the final result for the scalar
tilt
\begin{equation}
n_s \sim 1-7\varepsilon,
\end{equation}
which can be seen to arise as the limiting case of the $W \sim 1$ solution.
Using the WMAP bound, this serves to fix a much tighter bound on $\varepsilon$ at horizon crossing,
which we can interpret as a upper bound on $N$
\begin{equation}
N \le \frac{V_2 10^{2}M_p^2 h^4}{V_0},
\end{equation}
and therefore we can satisfy the experimental bounds on inflation.
In both cases we see that the indices are positive definite, 
as in the Ads models.
\end{itemize}

\subsection{Examples of other solutions}

Let us consider more general solutions which arise from the master equation
(\ref{eq:master}). We will consider both UV and IR inflationary solutions for generality
and then make some comments about the general signals of inflation in the large $N$ limit.

To distinguish between each solution branch,
we note that on the UV side we have large
field inflation. So there will generally be positive contributions to the potential,
and the vacuum energy $V_0$ will be set to zero. In the IR branch, we
have $V_0 \ne 0$ at all times and the potential will generally
be taken to be tachyonic indicating that the branes move away from the tip
of the throat.
In the fuzzy sphere picture, the UV side corresponds to a collapsing sphere,
while the IR side corresponds to an expanding sphere.

An interesting solution was analysed in \cite{tong} when there is no
quadratic term in the potential ie $V(\phi) = V_0 - V_4 \phi^4$.
Solving the master equation (\ref{eq:master}) for $H(\phi)$ and integrating
back to find the time dependence of the inflaton, we obtain the following
solutions for the AdS type backgrounds:
\begin{eqnarray}
H(\phi) &\sim& H_0 + H_4 \phi^4 \nonumber \\
\phi(t) &\sim& \frac{1}{4M_p \sqrt{(t_f-t)}} \sqrt{\frac{NW_0}{H_4}},
\end{eqnarray}
where the terms in the Hubble parameter are calculated to be
\begin{equation}
H_0 = \sqrt{\frac{NT_3 V_0}{3M_p^2}}, \hspace{0.5cm} H_4
= \sqrt{\frac{NT_3}{3V_0}}\frac{1}{2M_p} \left(\frac{W_0 -1}{L^4} \pm V_4 \right).
\end{equation}
Note that there is a potential sign ambiguity in the definition of $H_4$.
This is because we can consider either the IR inflation (where we have a minus
sign in the potential) or the UV inflation (where we have a plus sign).
In both cases, the equation of motion for the inflaton is the same.
In the IR case this is the early time, small $\phi$ solution,
whereas in the UV case it corresponds to the \emph{late} time solution where
the fuzzy sphere has collapsed to almost zero size. In both cases,
the form of the equation of motion implies that in the small field limit
the inflaton is moving relativistically with large $\gamma$.
The number of e-foldings can be determined as follows:
\begin{equation}
N_e \sim H_0(t_e-t_0) + \frac{NW_0}{16 M_p^2} \left(\phi^2(t_e)-\phi^2(t_0) \right),
\end{equation}
where $t_0$ and $t_e$ are the times at the beginning and end of inflation,
respectively. 
We must ensure that $t_f \ge t_e$, where $t_f$ represents the time at which we can no longer
trust our approximations.

Let us now reconsider the mass gap solution.
Since the throat is finite and the warp factor is actually constant for small
values of $\phi$, there is no reason why the field is moving rapidly near the
origin. Let us take a quadratic potential of the form
$V(\phi) = V_0 \pm V_2 \phi^2$, where the positive sign signifies UV type inflation.
For small values of the inflaton, we have the Hubble parameter
$H(\phi) = H_0 + H_2 \phi^2$, where the coefficients are determined
in a similar way to those in the AdS case.
If we assume that the field is small (in both cases), the general
solution can be written as
\begin{equation}
H_0 \sim \sqrt{\frac{NT_3 V_0}{3M_p^2}}, \hspace{0.5cm} H_2 \sim
\frac{3H_0}{8M_p^2} \left(1 \pm \sqrt{1+\frac{8NT_3 V_2}{9H_0^2}} \right).
\end{equation}
This solution is valid for the IR inflation. For the UV solution, one only
need substitute a minus sign in front of the $V_2$ term inside the square root.
Again this means that for the IR solution this is the early time evolution,
while for the UV solution it is the late time evolution of the Hubble parameter.
In both cases, the solution for the field can be written as
\begin{equation}
\phi(t) \sim \phi_0 e^{-\frac{4M_p^2 H_2 (t_{f}-t)}{N}}, \hspace{1cm} t_f \ge t,
\end{equation}
which means that the field is rolling non-relativistically
because $H_2$ and $\phi$ are both small.

Let us focus initially on the IR solution.
If we wish the inflaton to be increasing, corresponding to the branes moving
away from the origin, we are forced to choose the minus sign
in $H_2$. The other sign is a solution where the field is getting smaller.
In the event that $V_2$ is zero, we find either a solution where the inflaton
is at a constant value, or $H_2$ is positive definite and again the field
is getting smaller. The former case corresponds to de-Sitter expansion since
we can immediately integrate the solution for the scale factor to get
$a = e^{H_0 t}$.
In the UV region we must also impose an additional reality constraint
$9H_0^2 \ge 8NT_3 V_2$.
If this reality bound is saturated, the field is again rolling towards
the origin. In both cases the evolution of the field is essentially
determined by the vacuum energy $V_0$ which sets the overall scale for
the Hubble parameter.

Using the solution to the equation of motion, we can calculate the number
of e-foldings and write is as a function of the field.
In the UV case the branes would be near the bottom of the throat,
and so this correspond to late time evolution of the inflaton.
As such we interpret this as the early stage evolution of the
IR inflationary model. We obtain
\begin{equation}
N_e \sim H_0 (t_e-t_0) + \frac{N}{8M_p^2} \left(\phi^2(t_e)-\phi^2(t_0) \right),
\end{equation}
e-foldings in this regime and we must again ensure that $t_e \le t_f$.
Since we expect $\phi$ to be small over this time period in accordance with
our approximation of the Hubble parameter, the dominant contribution to the
number of e-foldings comes from the constant part of the potential.
However since $N >> 1$ we may well see
a sizable contribution to the number of e-foldings.

Let us consider what happens in more generality, although we will assume that the Hubble
parameter is generically of the form shown in (\ref{eq:largeNhubble}).
Focusing our attention on the power spectra, we find that the amplitudes are given by
\begin{eqnarray}
A_T^2 &=& \frac{2 H^2}{\pi^2 M_p^2}, \\
A_S^2 &=& \frac{H^2}{8\pi^2 M_p^2 \varepsilon} \left(1 +
\frac{4M_p^4 H'^2}{N^2 W^2 h^4 T_3} \right)^{1/2}, \nonumber \\
r&=& 16 \varepsilon \left(1+\frac{4M_P^4 H'^2}{N^2 W^2 h^4 T_3} \right)^{-1/2},
\nonumber
\end{eqnarray}
where we have included $r$ as the ratio of the tensor/curvature amplitudes.
Recall that each of these is to be evaluated at horizon crossing if we wish
to normalise them to the WMAP data~\cite{wmap}. Note that $\varepsilon$
is expected to be small at horizon crossing.
We can repeat the same analysis as before and consider limits of the
term in parenthesis.
(i) In the first case when we consider relativistic motion,
the scalar amplitude and ratio reduce to
\begin{eqnarray}
A_s^2 &\sim& \frac{H^2 |H'|}{4 \pi^2 \varepsilon N W h^2 \sqrt{T_3}} \nonumber \\
r&\sim& \frac{8 \varepsilon NW h^2 \sqrt{T_3}}{M_p^2 |H'|}.
\end{eqnarray}
If we satisfy the condition that $r \le 0.24$ then we find from the scalar amplitude, saturating the bound,
that $8 H^2 \sim 10^{-9} M_p^2$ or more concretely that
\begin{equation}
N V(\phi) \sim 10^{-10} \left(\frac{M_p}{M_s} \right)^4,
\end{equation}
assuming that the string coupling is around $10^{-2}$. This will be extremely difficult to 
satisfy under the assumption of large $N$, and also vacuum energy dominance, unless we are willing
to postulate a low string scale

(ii) If we consider the non-relativistic limit, we find the solutions
\begin{eqnarray}
A_s^2 &\sim& \frac{H^2}{8 \pi^2 M_p^2 \varepsilon}
\left(1 + \frac{2 M_p^2 H'^2}{N^2 W^2 h^4 T_3} + \ldots \right), \nonumber \\
r &\sim& 16 \varepsilon \left(1 - \frac{M_p^2 H'^2}{N^2 W^2 h^4 T_3}
+ \ldots \right).
\end{eqnarray}
The first expression implies that $\varepsilon >> H^2$ at horizon crossing,
which can be satisfied provided that the energy density of the inflaton
is vanishingly small. In fact if this condition is met, the ratio will
simultaneously be satisfied. A quick calculation shows that if we demand $r \le 0.24$ (ie using the strongest possible WMAP bound),
then the energy density must satisfy $E \le 10^{-7} M_P^4$ which is very small in Planck units.

Using the fact that $f_{NL} \le 100$ we can obtain a bound on $N$ for our theory. For an arbitrary warp factor we see that
\begin{equation}
N \ge \sqrt{\frac{32 \pi g_s}{299}} \frac{|H'|}{W h^2} \left(\frac{M_p}{M_s} \right)^2.
\end{equation}
To an order of magnitude approximation the numerical factor is $\mathcal{O}(1)$.
Now for mass gap backgrounds we may generally expect $W\sim 1$ or $W>>1$ and $h$ is a small constant which forces $N$ to be large. 
This will be further enhanced by a smaller string scale unless the derivative of the Hubble parameter is vanishingly small.
For the Ads scenarios we have competition between the $W$ and $h^2$ terms in the denominator - so we would expect $N$ to be
set explicitly by the choice of background fluxes and the string scale.

The general form for the scalar spectral index can be calculated to give
\begin{eqnarray}
n_S - 1 &=& -2\varepsilon \left(2 - \frac{H}{H'} \Delta \right) \\
\Delta &=& \frac{H''(\gamma^2+1)}{2H' \gamma^2} + \frac{(W^2-1)(2\gamma^2-1)}{W\gamma^2 \phi} 
+ \frac{h'}{h\gamma^2 W}\left(W(\gamma^2-1) + (W^2-1)(1-2\gamma^2)\right)  \nonumber
\end{eqnarray}
for an arbitrary warp factor. In principle the backreaction effects will appear through a redefinition of the harmonic function,
and so this expression should be valid for all theories satisfying our assumptions.
This equation simplifies once we assume relativistic motion, ie $\gamma >> 1$
\begin{equation}
\Delta \sim \frac{H''}{2H'} + \frac{2(W^2-1)}{W\phi} + \frac{h'}{Wh} \left(W - 2(W^2-1) \right).
\end{equation}
For AdS type solutions the fuzzy potential is constant, however we can see that the limiting solutions are
\begin{eqnarray}
\Delta &\sim& \frac{H''}{2H'} + \frac{h'}{h} \hspace{4cm} W \sim 1 \nonumber \\
\Delta &\sim& \frac{H''}{2H'} + 2W \left(\frac{1}{\phi}-\frac{h'}{h} \right)\hspace{1.7cm} W >>1.  
\end{eqnarray}
Clearly in our simplest case analyzed in the previous section we see that for $W>>1$ the second term will be identically zero
thus cancelling out any dangerous $1/\phi$ dependence. For the mass gap solution we find similar expressions to those presented above
except that the $h'$ terms will be zero - at least to leading order. Unless the backreaction dramatically alters the solution, we should
expect inflation to favour the $W>>1$ regime, in which case the scalar index is essentially only a function of the potential and its
derivatives - the overall scale being set by the $\varepsilon$ term.

In general we expect the number of e-foldings to be enhanced by factors of $N$ thus making the universe generically very flat. However
the factors of $N$ tend to increase the level of scalar and tensor perturbations, making it difficult to satisfy observational bounds without
imposing restrictive fine tuning.
\section{Inflation at finite $N$}

In this section, we investigate cosmic inflation due to a small number of
coincident branes. We begin with some general remarks about the
finite $N$ formalism before specialising to two simple cases, namely
$N=2$ and $N=3$.
In these cases the action is highly non-linear and gives
an expression for the speed of sound and the inflationary parameters in
certain regions of the phase space. They are very different from those in
the single brane models.

\subsection{General remarks and motivations}

We will now switch to the finite $N$ formulation of the non-Abelian Myers action,
using the prescription for the symmetrised trace as given in \cite{strconjecture}.
We believe this prescription to be correct, however a concrete proof remains an outstanding problem.
In most models of brane inflation, the bulk fluxes are tuned so that only
a single brane is left after the brane-flux annihilation process. In the context
of the landscape, this is a very special case and the general expectation is
that there remain several residual branes which will tend to coincide to
minimise their energy in a warped throat.
In the first part of the paper we looked at the large $N$ limit, which has
many problems due to the large back-reactive effects on the geometry
although it is a more general solution than the single brane cases.
In the remainder of the paper we will look at the solution when there
are a handful of residual branes at the tip of a warped throat. This means
that we can effectively neglect back-reactive effects as in the single brane
models, while still retaining the enhanced non-Abelian world-volume symmetry.

Let us rewrite the expressions for the energy and pressure of the coincident
branes in a more suitable manner for finite $N$:
\begin{eqnarray}\label{eq:finiteNenergy}
E &=& T_3 {\rm STr} \left(h^4 \sum_{k,p=0}^{\infty} (-X \dot{R}^2)^k Y^p
(\alpha^i \alpha^i)^{k+p}(1-2k)\left(\frac{1/2}{k}\right)\left(\frac{1/2}{p}
\right) + V - h^4 \right) \nonumber \\
P&=&-T_3 {\rm STr} \left(h^4 \sum_{k,p=0}^{\infty} (-X\dot{R}^2)^k Y^p
(\alpha^i \alpha^i)^{k+p} \left(\frac{1/2}{k} \right) \left(\frac{1/2}{p} \right)
+ V -h^4 \right),
\end{eqnarray}
where we have used the following definitions
\begin{equation}
X = \lambda^2 h^{-4}, \hspace{1cm} Y = 4 \lambda^2 R^4 h^{-4}, \hspace{1cm}
\left(\frac{1/2}{m} \right)= \frac{\Gamma(3/2)}{\Gamma(3/2-m)\Gamma(m+1)},
\end{equation}
and the fact that the potential is a singlet under the trace.
We employ the symmetrisation procedure in \cite{strconjecture}.
The basic formulas we need are then
\begin{eqnarray}\label{eq:prescription}
{\rm STr} [(\alpha^i \alpha^i)^m] &=&
2(2m+1) \sum_{i=1}^{(n+1)/2}(2i-1)^m \nonumber\\
&=& 2(2m+1) \sum_{i=1}^{n/2} (2i)^m .
\end{eqnarray}
The first line corresponds to odd $n = N-1$, and the second line to even $n$.
Note that we move from working with the $N$-dimensional representation
to the spin representation with $n=2J$. It is important to consider what
we mean by a physical radius in this context. We use the definition
\begin{equation}
r^2 = \lambda^2 R^2 {\rm Lim}_{m \to \infty} \left(\frac{STr
 (\alpha^i \alpha^i)^{m+1}}{STr (\alpha^j \alpha^j)^m} \right) = \lambda^2 R^2 n^2,
\end{equation}
which implies that the Lagrangian will converge for velocities from $0$ to $1$,
and moreover that the radius of this convergence will be unity.
This definition is consistent with what we know about the solution in the large
$N$ limit.

To illustrate the additional complexity arising from the finite N solution,
let us calculate the speed of sound in two examples using (\ref{eq:speedofsound}). For $N=2$, we find
\begin{equation}
C_s^2(N=2) = \frac{(1-X\dot{R}^2)(3+4Y-X\dot{R}^2[2+3Y])}{3+Y(4-X\dot{R}^2)},
\end{equation}
where $R^2 = r^2/\lambda^2$. This is obviously far more complicated than the
large $N$ expression which is $C_s^2 = 1 - X \dot{R}^2$.
For $N=3$, we obtain
\begin{equation}
C_s^2(N=3) = \frac{(1-4X\dot{R}^2)(3+16Y-8X\dot{R}^2[1+6Y])}{3+16Y(1-X\dot{R}^2)},
\end{equation}
where now $R^2 = r^2/(4 \lambda^2)$.
Note that in both cases, we recover the usual result that $C_s^2 =1$ when
the velocity of the branes is zero.

We may have anticipated the fact that at large $N$ the solution should resemble that of a single brane, through
knowledge of the Myers effect. However the finite $N$ solutions would be expected to be radically different.
One could imagine trying to start with the supergravity dual of this model, i.e some fivebrane wrapping a two-cycle
with $N$ unnits of $U(1)$ flux. However in order to capture finite $N$ corrections to the theory requires the use of the 
worldvolume star-product, and not simply the ordinary product assumed here. Extrapolating backwards to extremely small values
of $N$ requires us to keep higher order terms in the expansion of the star product, which significantly complicates the 
form of the resulting Lagrangian. 

\subsection{Two brane inflation.}

In this subsection, we consider inflation driven by two coincident branes
moving in the warped throat. This gives rise to a $U(2)$ symmetry on the
world-volume. The relevant expressions for the
energy and pressure can be calculated using (\ref{eq:finiteNenergy})
and (\ref{eq:prescription}):
\begin{eqnarray}\label{eq:N=2energy}
E &=& 2T_3 \left(\frac{h^4(1+2Y-XY\dot{R}^2)}{\sqrt{1+Y}(1-X\dot{R}^2)^{3/2}}
+V -h^4 \right) \\
P&=& -2T_3 \left(\frac{h^4(1+2Y-X\dot{R}^2[2+3Y])}{\sqrt{1+Y}\sqrt{1-X\dot{R}^2}}
+ V - h^4 \right). \nonumber
\end{eqnarray}
Note that in order to keep the energy finite, we should impose the constraint
$\dot{\phi}^2 \le h^4 T_3$. This also ensures that the contribution
coming from the DBI part of the action will be non-negative, as can be
seen from the first term in the numerator of the energy equation.

In general it is difficult to get solutions due to the complicated form of
the energy density. So let us make the approximation that the inflaton is
rolling ultra relativistically.
We can define the relativistic factor $\gamma$ much as we did in the large
$N$ solution by $\gamma = (1-X\dot{R}^2)^{-1/2}$.

We now write the energy and pressure as functions of $\gamma$, and then
take the large $\gamma$ limit. By utilising the conservation equation and
dropping all acceleration terms, we can find the solution
\begin{equation}\label{eq:gamma3}
\gamma^3 \sim \mp \frac{ M_p^2 H'}{\sqrt{T_3(1+Y)}h^2},
\end{equation}
where $H'$ is the derivative of the Hubble parameter with respect to $\phi$.
The sign in (\ref{eq:gamma3}) corresponds to the two choices
$\dot{\phi} = \pm \sqrt{T_3} h^2+... $ in the expansion of the velocity
$\dot{\phi}$ about its saturation value.
The $-$ sign is for the choice $\dot{\phi} > 0$ while the + sign for
$\dot{\phi} <0$. Note that in order to have $\gamma >0$,
we must demand that $H'<0 $ for $\dot{\phi} >0 $ and $H'>0$ for $\dot{\phi} <0$.
The choice of sign here is vital to obtaining the correct solution branch
for inflation.

We can rewrite the speed of sound as a function of $\gamma$ and $Y$ as follows:
\begin{equation}
C_s^2 = \frac{1}{\gamma^2} \left(\frac{\gamma^2(1+Y)+2+3Y}{3 \gamma^2(1+Y)
+Y}\right),
\end{equation}
which shows the finite $N$ corrections to the equation in this
limit. Without recourse to a specific background, we can make the
following observation: If we consider limiting solutions for $Y$, ie that it is either $>> 1$ or $<<1$, all
$Y$ dependence drops out of the expression and the equation
reduces to $C_s^2 \sim 1/(3 \gamma^2)$. The sound speed is thus
only a third of that in the large $N$ limit. This appears to be the
attractor point for the velocity with this action regardless of
background choice.
In this case we find (\ref{lambda}) is given by
\begin{eqnarray}
\Lambda =
\frac{X\dot{R}^2(5+6Y-XY\dot{R}^2)}{2(1-X\dot{R}^2)(3+4Y-XY\dot{R}^2)}\,.
\end{eqnarray}
In the $\gamma \gg 1$ case, the non-linearity parameter is
\begin{eqnarray}
f_{NL} \approx 0.24 \gamma ^2,
\end{eqnarray}
which is a little smaller than the large $N$ (and single brane) solution and effectively means
that we can satisfy the observational bounds whilst considering larger velocities than in the 
single brane scenario.

The corresponding Friedmann equation in this case is
\begin{equation}
H^2 = \frac{E}{3 M_p^2},
\end{equation}
where we are using the energy as defined in (\ref{eq:N=2energy}).
Substitute our expression for $\gamma$ into this equation,
we find that the Hamilton Jacobi equation for $N=2$ (in the large $\gamma$ limit)
becomes
\begin{equation}
H^2(\phi) \sim \frac{2 T_3}{3 M_p^2} \left( V(\phi) - h^4
\mp \frac{h^2 M_p^2 H'}{\sqrt{T_3}} \right),
\end{equation}
for an arbitrary flux induced potential.
At this stage, we could either specify the form of the Hubble parameter and
then consider how this modifies the potential \cite{uvmodels},
or we could specify the form of the potential and then solve for $H$.
We use this latter approach as this appears to be more within the spirit of
the Hamilton-Jacobi formalism we have employed thus far.

\subsubsection{Inflation in AdS type backgrounds}

Let us first consider a solution where the potential is dominated by
a constant, $V \sim V_0$.
Let us also assume that the background is approximately AdS.
In the small field limit, we expect the $h^4$ term is negligible compared to
the remaining terms. So as a first approximation, we ignore its contribution.
Solving this differential equation, with an appropriate constant of integration
$\tilde C$, we obtain
\begin{equation}\label{eq:n=2master}
H^2(\phi) \sim \frac{2T_3 V_0}{3M_p^2} {\rm tanh}^2 \left(
\sqrt{\frac{3V_0}{2M_p^2}} \frac{L^2}{\phi}(-1 + \tilde C \phi) \right).
\end{equation}
Since $H' <0 $, we have assumed $\dot{\phi} >0 $ in obtaining this solution.
In the limit of very small $\phi$, we can approximate the solution by
$H^2 \sim 2T_3 V_0 / (3 M_p^2)$, which is of the same functional
form as in the large $N$ case. However this expression is not consistent
with the our approximation that $\gamma >>1$ since the ratio $H'/\phi^2$ is
approximately zero for vanishingly small values of $\phi$.
Therefore we must be careful to choose a regime of validity where
this solution is valid. Careful inspection shows that the function
$\gamma^3$ has a turning point in the small field limit, with a maximal
value given by
\begin{equation}
\gamma^3_{\rm{max}} \sim \frac{256 e^{-4}}{9}\frac{M_p^4}{L^4 V_0^{3/2} \sqrt{1+Y}},
\end{equation}
where $Y$ is constant in the AdS background, and the value of the field
at this point is given by $\phi = \phi_{\rm{max}} \sim L^2\sqrt{6V_0}/(4M_p)$.
Thus to consider inflation in this region, we need some 'extreme fine tuning'
to set up the initial value of the field.

Rather than proceeding this way, we make the Taylor series expansion of
the Hubble parameter for small $\phi$ but without dropping
the $h^4$ term in (\ref{eq:n=2master}). This means we must include quartic
terms in the expansion of the Hubble parameter and so we need
\begin{equation}
H(\phi) \sim \sum_{i=0}^4 H_i \phi^i ,
\end{equation}
We will also keep quartic terms in the inflationary potential for consistency,
$V(\phi) \sim V_0 - V_2 \phi^2/2 - V_4 \phi^4/4 + \ldots$.
Equating the various coefficients, we find that the linear term in the Hubble
parameter actually vanishes, leaving us with the residual terms
\begin{eqnarray}
H_0 &=& \sqrt{\frac{2T_3 V_0}{3M_p^2}}, \\
H_2 &=& -\frac{T_3 V_2}{6 M_p^2 H_0}, \nonumber \\
H_3 &=& \frac{ 2 \sqrt{T_3} H_2}{3 L^2 H_0}, \nonumber \\
H_4 &=& -\frac{1}{12 M_p^2 L^4 H_0 \sqrt{T_3}} \left(V_4 L^4 T_3 ^{3/2}
+ 4 T_3^{3/2} - 12 M_p^2 H_3 L^2 T_3 + 6M_p^2 L^4 \sqrt{T_3}H_2^2 \right).\nonumber
\end{eqnarray}
We find that the constant piece of the potential dominates the Hubble
parameter when the field is vanishingly small.
The sign of the last term is potentially ambiguous which can lead to
interesting cosmological behaviour. It turns out that the Hubble term is
extremised at the usual $\phi=0$ solution (which is a local maximum),
and there exists a non-trivial solution given by
\begin{equation}
\phi_{min} = \frac{1}{8 H_4} \left(-3 H_3 \pm \sqrt{9H_3^2-32 H_2 H_4} \right),
\end{equation}
where we must require the term inside the square root to be non-negative.
We can rewrite this reality constraint as
\begin{equation}
\frac{H_2}{H_4} \ge \frac{8 H_0^2 L^4}{T_3}.
\end{equation}
We now find the possibility of a 'cosmic turnaround' because $H_2$ will be
negative definite for those regions of phase space where $H_4$ is also negative.
If we concentrate on regions of $\phi $ near the origin, $H' <0 $ so we
necessarily have $\dot{\phi}> 0 $ in order for $\gamma >0 $ in our approximation
(\ref{eq:gamma3}). Of course, we have implicitly assumed that the field is monotonic
so we cannot say anything about the reality of such a bounce solution within the
current framework.

It is easy to solve the equation of motion in this limit for relativistic motion.
As in the case of a single brane, and for the large $N$ solution,
we obtain the following term for the inflaton equation of motion
\begin{equation}
\phi \sim \frac{\phi_0}{1 - \phi_0\sqrt{T_3} (t-t_0)/L^2},
\end{equation}
where again we define $\phi_0$ as the field value at time $t=t_0$,
and it can be seen that $\dot{\phi } >0 $.

To see the implications of this for inflation, we must first determine
which are the relevant parameters in this finite $N$ formulation.
The modified 'fast roll' parameter in this case can be written as
\begin{equation}\label{eq:slowroll}
\varepsilon \sim \pm \frac{\phi^2 \sqrt{T_3}}{L^2} \frac{H'}{H^2},
\end{equation}
where the sign is related to the sign for $\dot{\phi}$ and we need to demand
that inflation ends when $\varepsilon=1$ as usual.
Recall that for small $\phi$, $H' <0 $ and we need to choose the minus sign
in this equation (coming from the choice $\dot{\phi}>0 $).
Inserting our expression into the Hubble parameter, we see that
$\varepsilon$ can be expanded in powers of the inflaton.
Keeping only the leading order term (which amounts to dropping
$\mathcal O(\phi^4)$ contributions), we see that inflation will end around
\begin{equation}
\phi_e \sim \left(\frac{L^2 H_0^2}{2 |H_2| \sqrt{T_3}} \right)^{1/3}.
\end{equation}
The corresponding number of e-foldings given by this Hubble parameter is
generically a power series in $\phi$. We expect the dominant contribution
to arise from the constant piece $H_0$ as in the standard inflationary scenario.
Integrating over the field, we find the expressions for the inflaton as
a function of e-folding number:
\begin{equation}
\phi_0 \sim \phi_e \left(1+\frac{N_e \sqrt{T_3} \phi_e}{L^2 H_0} \right)^{-1}.
\end{equation}
Inserting this back into the 'fast roll' parameter (\ref{eq:slowroll}),
we find that the dependence on the number of e-foldings is
of the same functional form as in the large $N$ case:
\begin{equation}
\varepsilon \sim \left(1 + N_e \left(\frac{3M^2_p}{L^4V_2} \right)^{1/3}\right)^{-3}
\end{equation}
For the perturbation amplitudes we can use the general results developed in the appendix. We find that the gravitational wave amplitude will
be constant to leading order, and given by
\begin{equation}\label{eq:n=2tensor}
\mathcal{A}_T^2 \sim \frac{4 T_3 V_0}{3 \pi^2 M_p^2} = \frac{V_0}{6 \pi^5 g_s} \left( \frac{M_s}{M_p}\right)^4
\end{equation}
the corresponding expression for the scalar amplitude is given by
\begin{equation}
\mathcal{A}_S^2 \sim \frac{V_0 \gamma}{32 \sqrt{3} \pi^5 g_s} \left(\frac{M_s}{M_p} \right)^4\left(1 + 60 \left\lbrace\frac{3M_p^2}{L^4 V_2} \right\rbrace^{1/3} \right)^3
\end{equation}
We choose to re-write this in terms of the tensor amplitude as
\begin{equation}\label{eq:n=2scalaramp}
\mathcal{A}_S^2 \sim 10^{-1}\gamma \mathcal{A}^2_T \left(1 + 60 \left\lbrace \frac{3M_p^2}{L^4 V_2} \right\rbrace^{1/3} \right)^3
\end{equation}
Now the non-gaussianity condition implies that the bound $1 << \gamma^2 << 400$ must be satisfied to comply with observation.
The spectral indices for this model at large $\gamma$ are
\begin{eqnarray}\label{eq:n=2indices}
n_T &\sim& \frac{2 H'\sqrt{T_3} \phi^2}{H^2 L^2}\\
n_S - 1 &\sim& -\frac{4X_2}{(1+60X_2)} + \ldots \nonumber
\end{eqnarray}
where the first line is understood to be evaluated at horizon crossing, and we have written $X_2 = (3 M_p^2/ L^4 V_2)^{1/3}$ as a dimensionless parameter. Note that the tensor index is negative, but suppressed by the Hubble parameter.
Clearly the scalar index is bounded from above by unity, and so normalising to WMAP data implies that $0 \le X_2 \le 0.05$, or more concretely that
\begin{equation}\label{eq:n=2constraint}
\frac{L^4 V_2}{2.4 \times 10^{4}} \ge M_p^2.
\end{equation}
Using this constraint in (\ref{eq:n=2scalaramp}) we see that the term in brackets varies between $1$ and $64$. The relationship between the 
two amplitudes is characterised by the parameter $r$ and so we recover the anticipated DBI relation
\begin{equation}
r \sim \frac{1}{\gamma}.
\end{equation}
The fact that $ 1 << \gamma \le 20$ in this model implies that $r$ will generically be small and thus well within the WMAP confidence bounds.
Furthermore this implies that the tensor amplitude will be smaller in magnitude than the scalar one, something like $10^{-10}$ for a range of
$\gamma$.
Using the normalisation for the scalar amplitude, namely that it satisfies $\mathcal{A}_S^2 \sim 10^{-9}$ at horizon crossing, we see that
this constrains the potential in terms of the string scale. For small $X_2$ we find
\begin{equation}
\frac{V_0 \mathcal{O}(10^3)}{g_s} \left(\frac{M_s}{M_p} \right)^4 \ge 1
\end{equation}
where we are interested in an order of magnitude approximation. Whilst for the maximal value of $X_2$ we recover the same constraint but with an additional
factor of $10^3$ in the numerator. Clearly both solutions are sensitive to the mass splitting between Planck and string scales and imply
that generic inflation prefers the string scale to be close to the Planck scale in order not to have super-Planckian scalar potentials. For 
example if we have $M_S \sim 10^{-1} M_p$ then the potential constraint becomes $\mathcal{O}(10^1 - 10^4) V_0 \ge 1$.
What about the constraint in (\ref{eq:n=2constraint})? Upon substituting for the background parameters we see that this equation can
be written as
\begin{equation}
V_2 \ge \frac{\mathcal{O}(10^5)}{MK} \frac{M_p^2}{M_s^4}
\end{equation}
which sets the inflaton mass scale.

\subsubsection{Inflation in mass gap backgrounds}

If we repeat the analysis for the mass gap backgrounds,
(assuming that the constant part of the potential dominates),
we find the following solution for the Hubble parameter:
\begin{equation}
H(\phi) =\pm \sqrt{\frac{2T_3}{2M_p^2}\left(V_0 - h^4 \right)}\;
\tanh \left(\sqrt{\frac{3(V_0-h^4)}{2 M_p^2}}\frac{\phi + \tilde C}{h^2} \right)
\end{equation}
where we have used the fact that the warp factor is constant to write the
solution as a function of $h$.

Substituting this into the gamma constraint, we must require that the solution
is larger than unity even when $\phi$ is vanishingly small.
We make a Taylor series expansion of the resultant function, and find that
$\gamma^3 \propto {\rm sech}^2( \rm{F}(\tilde C))$, where the amplitude of
the function is determined by the ratio of the potential and the warp factor.
Also note that $F(\tilde C)$ is an function of the Casimir, whose precise form is not relevant 
for understanding the physics of the solution.
Now the hyperbolic trigonometric function is a decreasing function
of its argument, which forces us to take the limit $\tilde C^2 << h^4/V_0$
in order for the large velocity expansion to hold.

Let us assume that we can in fact take this limit
and consider the implications for inflation.
Calculation of the fast roll parameter $\varepsilon$ yields the following
\begin{equation}
\varepsilon \sim \frac{3}{2} \left(\rm{Cosh}^2 \left(
\frac{\sqrt{6(V_0-h^4)}\phi}{2M_p h^2} \right) -1 \right)^{-1},
\end{equation}
which is a decreasing function of the inflaton field. Thus after some critical
field value $\phi_c$, we will find a solution where inflation
never ends.
It may appear that this is an artifact due to the neglect of higher order
terms in the potential. However if we consider quartic terms in $V(\phi)$,
and also up to the same order in a Taylor expansion of $H(\phi)$,
we find the same result that $\varepsilon$ is a decreasing
function of the inflaton. We conclude that in the relativistic limit that
inflation (once started) never ends\footnote{Technically this is no longer true
once the branes reach the gluing region. However the effective action is no
longer expected to be a good description of the physics in this region.} unless
we turn on extra effects, such as nonzero gauge fields.
This result is not anything that we expect from the results of the single
brane case \cite{tong}, or the large $N$ limit discussed in the previous sections
and appears to be a distinctly finite $N$ effect. Of course, it may well be
that standard inflation can occur for moderate values of $\gamma$.
However one should probably need numerical results to see it.

Of course, the mass gap background will eventually give way to something similar to the
Ads solution, where the harmonic function will have explicit dependence upon the inflaton field. 
Therefore we expect inflation to end in this regime. The fact that the mass gap solution has
finite warping means that it will be relatively easy to produce the necessary $60$ e-folds of expansion.
\subsubsection{Non-relativistic limit}
Let us restrict ourselves to the non-relativistic regime in order to see
the consequences for brane inflation \footnote{This has recently been examined in \cite{slowrollexpansion}.}.
It would of course be more preferable to obtain an interpolating solution between these two extremes, however it is analytically challenging
and would be better suited to a numerical analysis.
After performing a series expansion of the continuity equation,
we find the following solution for the velocity of the field in the Hamilton-Jacobi
formalism:
\begin{equation}
\dot{\phi} = -\frac{\sqrt{1+Y}M_p^2 H'}{(3+4Y)},
\end{equation}
which means that the corresponding Friedmann equation reduces to
\begin{equation}
\frac{3M_p^2 H^2}{2 T_3} \sim \frac{h^4}{\sqrt{1+Y}}\left(1+2Y
+\frac{Z_2 M_p^4 H'^2}{2 T_3 h^4} \right) +V(\phi) - h^4,
\end{equation}
where $Z_2 = (1+Y)/(3+4Y)$. Let us solve this equation by considering
a standard quadratic potential. There are two solution branches, one of which
has an imaginary component of $H$.
We ignore this solution as it appears unphysical. The other real solution
can be parameterised by a quadratic Hubble parameter with nonzero components
given by
\begin{equation}
H_0 = \sqrt{\frac{2T_3 V_0}{3M_p^2}} \hspace{1cm} H_2 = \frac{9H_0}{M_p^2}\left(1 \pm \sqrt{1 + \frac{V_2 M_p^2}{18 V_0}} \right).
\end{equation}
For inflation to occur we must ensure that we take the minus sign in the solution for $H_2$.

The speed of sound in this instance reduces to
\begin{equation}
C_s^2 \sim 1 - \frac{15 \dot{\phi}^2}{h^4 T_3(3+4Y)} \left(1+\frac{5Y}{4} + \ldots \right)
\end{equation}
where we can substitute $\dot{\phi}$ for derivatives of the Hubble parameter. Of course our non-relativistic expansion
assumes that the subleading term is smaller than unity in order for the speed of sound to remain a real function. This constrains
the field to satisfy
\begin{equation}
\dot{\phi}^2 << \frac{h^4 T_3 (3+4Y)}{15}.
\end{equation}
The non-relativistic assumption means that we can find inflationary solutions even in the mass gap
backgrounds. Using the definition of the $\varepsilon$ parameter we find that the leading order
contribution yields
\begin{equation}
\varepsilon \sim e^{2 \beta N_e}
\end{equation}
at $N_e$ e-folds before the end of inflation. We have introduced the dimensionless ratio $\beta = M_p^2 H_2/H_0$ for simplicity. For
inflation to occur we must ensure $\beta < 0$, or that $H_2 < 0$.
However the fact that the slow roll parameter is now exponential implies that the level of scalar perturbations will now be enhanced by
this exponential term. Our assumptions assume that $\phi$ is small, so the speed of sound is essentially unity in this instance.
It transpires that the simplest equation to study is the tensor to scalar ratio $r$, which is now given by $ r \sim 16 \varepsilon$.
If we demand that $r \le 1/4$ to satisfy the more stringent bound, then after some algebra we find the following constraint
\begin{equation}
V_2 \ge \frac{0.1 V_0}{M_p^2}
\end{equation}
where we have explicitly left in the numerical value. If this bound is not satisfied then we find $H_2$ to be very small 
which suppresses the number of e-foldings and sufficient inflation is generically difficult to achieve.


For Ads type backgrounds we have the following solution for the Hubble parameter
\begin{equation}
H_0 = \sqrt{\frac{2T_3V_0}{3 M_p^2}} \hspace{1cm} H_2 =\frac{3 H_0 (3+4Y)}{M_p^2 \sqrt{1+Y}}\left(1 \pm \sqrt{1+\frac{T_3 V_2 \sqrt{1+Y}}{9H_0^2(3+4Y)}} \right)
\end{equation}
where we must again take the minus sign for an inflationary solution. Interestingly for $Y<<1$ we see that the solution becomes exactly the 
same as the mass gap one.
We can follow the same procedure and obtain a similar result for the number of e-folds. The difference is of course due to the
constant nature of the factor $Y$.
\begin{equation}
\varepsilon \sim \exp \left(\frac{2 \delta N_e \sqrt{1+Y}}{3+4Y} \right)
\end{equation}
where $\delta$ is defined in a similar way to $\beta$ except that the Hubble parameters are different in this case.

The observed constraint on the amplitude ratio can now be written as
\begin{equation}
V_2 \ge \frac{0.1 V_0}{M_p^2} F(Y) \hspace{1cm} F(Y) = \frac{3+4Y}{\sqrt{1+Y}}
\end{equation}
which is slightly different from the mass gap solutions. The function $F(Y)$ acts to increase the rhs of the expression above, ranging from
$F(Y) \sim 3 \to 4 \sqrt{Y}$ depending upon our choice of fluxes. Larger values of $Y$ impose clearly impose tighter constraints on the parameter $V_2$
and so one would anticipate that smaller values are more preferential.

In both instances we see that in order to satisfy the scalar curvature constraints, we require the dominant term in the potential to
satisfy the following
\begin{equation}
V_0 \le 3 \times 10^{-7} \left(\frac{M_p}{M_s} \right)^4,
\end{equation}
which can be combined with the expressions for $V_2$ to yield a constraint purely on that variable in terms of the Planck and string scales
(and also the fluxes for the Ads case). We have again assumed a string coupling of $g_s \sim 10^{-2}$ in the above expression.
\subsection{Three brane inflation}
Let us now move to the case where there are three coincident branes,
giving rise to a $U(3)$ world-volume symmetry.
We have the energy and pressure:
\begin{eqnarray}\label{eq:N=3energy}
E &=& 2T_3 \left(\frac{h^4(1+4Y[1+X\dot{R}^2])}{\sqrt{1+2Y}(1-2X\dot{R}^2)^{3/2}}
+\frac{3V}{2}-\frac{3h^4}{2} \right), \\
P&=& -2T_3 \left(\frac{h^4(1+4Y-4X\dot{R}^2[1+3Y])}{\sqrt{1-2X\dot{R}^2}
\sqrt{1+2Y}} + \frac{3V}{2}-\frac{3h^4}{2} \right). \nonumber
\end{eqnarray}
The symmetry breaking induced by a gauge field in this case will be
$U(3) \to SU(3) \times U(1)$.
Let us again consider the large $\gamma$ solution for the fast rolling action.
It is convenient to define
\begin{equation}
\gamma = \frac{1}{\sqrt{1-2X\dot{R}^2}},
\end{equation}
which allows us to write the energy and pressure in (\ref{eq:N=3energy})
as explicit functions of $\gamma$.

However the exact solution for the speed of sound can be written as a function of $\gamma$
\begin{equation}
C_s^2 = \left(\frac{\gamma^2 -2}{\gamma^2}\right) \left(\frac{\gamma^2(1+8Y)
-4(1+6Y)}{\gamma^2(3+8Y)+8Y} \right),
\end{equation}
which, unlike the other solutions studied so far, allows for the possibility
that the sound speed is zero. This is the case if
either of the following critical conditions are satisfied:
\begin{equation}
\gamma_c^2 = 2 \hspace{1cm} \rm{or}, \hspace{1cm}
\gamma_c^2 = \frac{4(1+6Y)}{(1+8Y)}.
\end{equation}
The first condition corresponds to $\dot{\phi}^2 / h^4 = T_3/4$.
The second condition is a little more difficult to deal with due to
the potential $\phi$-dependence of $Y$.
Now we can consider the two simplifying limits.
(i) In the limit where $Y \to 0$, we see that the constraint on the velocity
becomes $\dot{\phi}^2/h^4 = 3 T_3/8$, while (ii) in the converse limit
(where $Y$ is dominant), we see that $\dot{\phi}^2/h^4 = T_3/3$.
All of these conditions are allowed because they satisfy the causality
constraint on the velocity. We know that fluctuation modes exit the horizon
at the reduced scale $kC_s = a H$ in these models, so a zero speed of sound
tells us that the modes will never exit the horizon.
In order to consider an inflationary epoch, we have to ensure the velocity is
either much smaller than either of the critical bounds (corresponding to
non-relativistic motion), or much higher corresponding to ultra relativistic
motion. Thus unlike the case of $N=2$, we are lead to selecting a specific
velocity range.
Even for $Y \sim \mathcal{O}(1)$, we find that $C_s$ rapidly tends towards unity as
in normal models of scalar field inflation.

In order to consider inflationary solution, we start with the continuity
equation. Taking the large velocity,
we find the general result
\begin{equation}\label{eq:N=3gamma}
\gamma^3 = \frac{- (\pm 1)M_p^2 H' \sqrt{2(1+2Y)}}{h^2 \sqrt{T_3}(1+6Y)},
\end{equation}
where we have made use of the fact that $\dot{\phi} = \pm \sqrt{2 T_3} h^2+..$
in this limit. The sign ambiguity here can be resolved by demanding
$\gamma$ to be positive. Since we are interested in solutions where $H' <0 $,
we take the + sign in the definition of the velocity.
Substituting our expression back into the Friedmann equation, using the formula (\ref{eq:N=3energy}) for the energy density,
yields the Hamilton-Jacobi equation
\begin{equation}
\frac{M_p^2 H^2}{T_3} = V(\phi) - h^4 - \frac{h^2 M_p^2 H'}{3}\sqrt{\frac{8}{T_3}},
\end{equation}
which can again be integrated to solve for $H$ once we specify
the background potential.
The level of non gaussianities arising from this action can be parameterised by
\begin{equation}
f_{NL} \sim \frac{162}{85(1+8Y)} \left(1 + \frac{10(1+8Y)}{51}\gamma^2 \right)
\end{equation}
which clearly has non-trivial dependence on the inflaton field for the mass gap backgrounds (since $Y$ is constant for the Ads solutions).
Let us explore the possible solution space here. For the Ads case we find that
\begin{equation}
Y = \frac{KM g_s}{4a\pi}
\end{equation}
and so can be small with appropriate tuning of the fluxes and the string coupling.
If we assume $Y<<1$ then we see that the non-gaussianities are (up to $\mathcal{O}(Y^2)$ terms - and dropping the constant piece)
\begin{equation}
f_{NL} \sim 0.37 \gamma^2 + \ldots,
\end{equation}
whilst if we assume that $Y$ is large (corresponding to large fluxes) we find the following
\begin{equation}
f_{NL} \sim 2.99 \gamma^2 + \mathcal{O}\left(\frac{1}{Y}\right).
\end{equation}
The latter condition is much larger than anything encountered before, and severely restricts the relativistic approximation
we have been making. In fact if we have $Y \sim \mathcal{O}(1)$ we find a similar condition.
However for small $Y$ we see that the non-gaussianities are roughly the same as in the previous 
sections, and would appear to be the more favourable regime for inflation. This tells us 
that we require $g_s << 1/(MK)$.

\subsubsection{Inflation in AdS type backgrounds}

It is generically difficult to find inflationary solution for AdS backgrounds.
To proceed with our small $\phi$, but large gamma solution, we again turn
to a Taylor series approach to the Hubble parameter.
Let us take the same form for the expansion as in the last section,
with a similar expression for the inflaton potential.
Again we find that there is no linear dependence in this limit,
but the non zero coefficients can be seen to be
\begin{eqnarray}
H_0 &=& \sqrt{\frac{V_0 T_3}{M_p^2}}, \\
H_2 &=& -\frac{V_2 T_3}{4 M_p^2 H_0}, \nonumber \\
H_3 &=& -\frac{\sqrt{8 T_3} H_2}{3 L^2 H_0}, \nonumber \\
H_4 &=& -\frac{1}{8M_p^2L^4H_0} \left( V_4 L^4 T_3 + 4T_3 + 8M_p^2 H_3
L^2 \sqrt{2 T_3} + 4M_p^2 L^4 H_2^2\right). \nonumber
\end{eqnarray}
The conditions for inflation are basically the same as in the previous section,
which is to be expected since we are assuming that
inflation is essentially driven by the constant part of the Hubble parameter.
The slow roll parameter is shifted only slightly by the extra brane because
the velocity in this case is increased by an extra factor of $\sqrt{2}$.
Solving for the inflaton at the end of inflation, we find
\begin{equation}
\phi_e \sim \left(\frac{L^2 H_0^2}{\sqrt{8 T_3} |H_2|} \right)^{1/3},
\end{equation}
where we have absorbed the minus sign into the definition of $|H_2|$
to make the solution manifestly positive. This is only slightly different from
that obtained in the $N=2$ case.
By integrating the Hubble term, we can invert again the resulting expression
to obtain the inflaton as a function of the number of
e-foldings. The result is the same as for $N=2$ expect now the tension
is doubled. Finally we obtain $\varepsilon$ as a function
of the number of e-foldings
\begin{equation}
\varepsilon \sim \left(1 + N_e \left(\frac{4 M_p^2}{L^4 V_2} \right)^{1/3} \right)^{-3}
\end{equation}
which represents only a slight numerical shift with regard to the expression
in the previous section for $N=2$.
We can once again calculate the relevant signals for this model, and
the analysis proceeds much as in the case of $N=2$, except that we are now
forced to restrict ourselves to the $\gamma >>1$ solution.
The tensor amplitude at leading order becomes
\begin{equation}
\mathcal{A}_T^2 \sim \frac{V_0}{4 \pi^4 g_s} \left(\frac{M_s}{M_p} \right)^4
\end{equation}
which is a factor of $3/2$ larger than the amplitude in the $N=2$ case (\ref{eq:n=2tensor}).
Whilst the scalar amplitude can again be written solely in terms of the tensor amplitude divided by the parameter $r$.

The tensor spectral index is relatively suppressed as in the $N=2$ case, however for the scalar index we find
\begin{equation}
n_S - 1 \sim -\frac{4X_3}{1+60X_3}
\end{equation}
which is identical in form to the $N=2$ solution in (\ref{eq:n=2indices}) under the replacement $X_3 = (4M_p^2/L^4 V_2)^{1/3}$. The 
same remarks apply here except the physical constraint is slightly tighter than before
\begin{equation}
\frac{L^4 V_2}{3.2 \times 10^4} \ge M_p^2.
\end{equation}
Using the small $Y$ constraint in order to suppress the non-gaussianities we can write this constraint purely in terms of the potential term
\begin{equation}
V_2 >> \frac{g_s \times 10^9 M_p^2}{M_s^4} \hspace{1cm} (Y<<1)
\end{equation}
Let us consider the two limiting solutions, bearing in mind that we expect that the $Y>>1$ case will lead to extremely large non-gaussianities.
For small $Y$ we see that the sound speed becomes
\begin{equation}
C_s(Y << 1) \sim \frac{\sqrt{\gamma^2 - 4}}{3 \gamma}
\end{equation}
and so if we also assume that $\gamma^2 >> 4$ then we see that this becomes $1/3$. This is unlike all the other DBI models studies so far.
Interestingly if we take the limit where $Y>>1$ we also see that it drops out of the analysis
\begin{equation}
C_s(Y >> 1) \sim \frac{\sqrt{\gamma^2-28/8}}{\gamma}
\end{equation}
and in fact we find that $C_s \sim 1$ as in standard slow roll models of inflation. Following the same procedure as in the 
$N=2$ case we can constrain the potential using the scalar amplitude. We find that the range for $V_0$ is
\begin{eqnarray}
V_0 &\sim& \mathcal{O}(10^{-10}-10^{-9})\left( \frac{M_p}{M_s}\right)^4 \hspace{1cm} Y << 1\\
V_0 &\sim& \mathcal{O}(10^{-9} - 10^{-7}) \left(\frac{M_p}{M_s} \right)^4 \hspace{1.15cm} Y >> 1 \nonumber
\end{eqnarray}
which once again indicates the sensitivity of inflation to the string scale.
\subsubsection{Inflation in mass gap backgrounds.}
Let us now restrict our analysis to the mass gap backgrounds.
We can again solve the master equation assuming
that the constant part of the potential dominates the solution. The result is
\begin{equation}
H(\phi) = \frac{\sqrt{T_3(V_0 - h^4)}}{M_p} \rm{tanh} \left(\frac{3(\phi
+ \tilde C)}{\sqrt{8} M_p h^2} \sqrt{V_0 - h^4} \right),
\end{equation}
which should be valid for small values of the inflaton field, and we have
left the mass gap warp factor as an arbitrary constant.
We must ensure that this expression is consistent with our demand that
the $\gamma$ factor is large. This requires us firstly to take the minus
sign in the velocity term, and secondly to examine the behaviour of the
function for small values of the inflaton.
Differentiating this function and then performing a Taylor series expansion
of (\ref{eq:N=3gamma}) for small $\phi$ yields the constraint
(valid up to terms of $\mathcal{O}(\phi^4)$)
\begin{equation}
3Q >> \left \lbrace1 + {\rm cosh}\left(\frac{3 \tilde C}{M_p} \sqrt{\frac{Q}{2}}
\right) \right\rbrace \left(1 + \frac{7 \phi^4 L^4}{4 \mu^4 \lambda^2 T_3^2}
\right),
\end{equation}
where we have introduced the simplifying notation $Q = V_0/h^4-1$.
Clearly to satisfy this condition, we must require that the term $\tilde C$
arising from the boundary condition be very small in Planck units.
Neglecting the $\phi^4$ terms, we can again use a Taylor series expansion,
this time for $\tilde C \sim 0$. This corresponds to a specific choice of the boundary conditions for the solution. 
In the leading order, we must satisfy
the following condition on the parameter $Q$:
\begin{equation}
Q >> \frac{2}{3}\left(1+\frac{3 \tilde C^2}{4 M_p^2} + \ldots \right).
\end{equation}
However the fact that we require $\tilde C$ to be small has an effect on
the amount of inflation we can have in this fast rolling regime.
To see this, let us calculate the fast rolling parameter $\varepsilon$,
making use of our near relativistic approximation. A short calculation
shows that
\begin{equation}
\varepsilon \sim -\frac{3}{2}{\rm csch}^2 \left(\frac{3(\phi
+ \tilde C)}{M_p}\sqrt{\frac{Q}{8}} \right).
\end{equation}
For small values of $\tilde C$ and the inflaton field,
we see that the real part of this function is divergent. In fact $\varepsilon$ is
a decreasing function of $\phi$ which suggests that inflation will only
be possible once the field reaches a critical point given by
$\phi_c \sim \frac{M_p}{3} \sqrt{\frac{8}{Q}} {\rm Arccsch}(\sqrt{\frac{2}{3}})
- \tilde C$, after which we enter a phase of eternal inflation which will not
end within the bounds set by our theory. Of course this may no longer be true once
higher order terms are included, and we leave this possibility for future work.

\subsubsection{Non-relativistic limits}
In this subsection, we examine the non-relativistic motion of the branes
and compare with the results from the previous sections.
>From the continuity equation and the definition of the energy momentum tensor,
we find the inflaton velocity
\begin{equation}
\dot{\phi} = -\frac{2M_p^2 H' \sqrt{1+2Y}}{3(1+4Y)}.
\end{equation}
Upon substitution of this back into the Friedmann equation, we obtain
\begin{equation}\label{eq:slowroll3}
\frac{3 M_p^2 H^2}{2 T_3} \sim \frac{h^4}{\sqrt{1+2Y}} \left(1 + 4Y
+ \frac{Z_3M_p^4 H'^2}{9 T_3 h^4} \right) + \frac{3}{2}(V(\phi) - h^4),
\end{equation}
where we have introduced another function $Z_3$:
\begin{equation}
Z_3 = \frac{(3+16Y) (1+2Y)}{(1+4Y)^2}.
\end{equation}
The above equation (\ref{eq:slowroll3}) is difficult to solve analytically
for either background. So we resort to the usual trick of Taylor expanding
the Hubble term for a given potential.

Let us consider the mass gap backgrounds. The simplest analytic solutions
are obtained when we keep only terms up to quadratic in the potential and
Hubble parameter. It is easy to see that the coefficient $H_1$ is imaginary
and so we drop it from the analysis. The results for the remaining components are
\begin{eqnarray}
H_0 &=& \sqrt{\frac{T_3}{2M_p^2}\left(3(V_0-h^4) + 2h^2 \right)}, \\
H_2 &=& \frac{9H_0 h^2}{8M_p^2} \left(1\mp \sqrt{1+\frac{4 V_2 T_3}{9h^2 H_0^2}}
\right).
\end{eqnarray}
The general result for the $\varepsilon$ equation reduces to
\begin{equation}
\varepsilon \sim \frac{8M_p^2 H_2 \phi^2}{3 H_0^2}\frac{\sqrt{1+2Y}}{(1+4Y)},
\end{equation}
for all backgrounds, where $Y$ is a function of the inflaton for
the mass gap solutions. For small values of $\phi$, we can expand this
and obtain a value for the field at the end of inflation.
As a result we can write the slow roll parameter as an explicit function of
the number of e-foldings:
\begin{equation}
\varepsilon \sim e^{\frac{8\beta}{3} N_e},
\end{equation}
where we have used the previous definition of $\beta$.
For any inflation to occur we must have $\beta < 0 \to H_2 < 0$ therefore we must
again choose the minus sign in the expression above.

Following the same line of reasoning as in the $N=2$ case we see that the constraint on the
potential contributions can be written as follows
\begin{equation}
V_2 \ge \frac{10^{-3}V_0}{M_p^2}
\end{equation}
where we have neglected higher order contributions in $h^2$.
This is smaller than the constraint in the two-brane solution due to the additional 'mass' coming from the
extra brane. The inflaton is effectively weighted by this contribution thus making it roll more slowly.
Of course this analysis is only valid for small velocities, which is more problematic in this instance as there
are zeros for the speed of sound function which destroys any hope of obtaining an inflationary solution.

We can also obtain a simple analytic solution if the constant parts of the
potential and the Hubble parameter are the dominant contribution.
In this case, we find
\begin{equation}
H_0 = \sqrt{\frac{T_3}{M_p^2} \left(V_0 - h^4 + \frac{2 h^2}{3} \right)},
\end{equation}
where the warp factor contribution is subdominant.
This implies that the velocity of the inflaton will be zero, as can be noted from the
Hamilton-Jacobi expression.

As for the AdS type backgrounds, we again cannot obtain simple analytic
solutions when we keep quartic terms in the Friedmann equation. So again we
restrict our analysis to the purely quadratic pieces. As in the other cases,
the linear term in $H$ must vanish for consistency, and so
the physical solutions are
\begin{eqnarray}
H_0 &=& \sqrt{\frac{3 T_3 V_0}{2 M_p^2}} \\
H_2 &=& \frac{H_0 \sqrt{1+2Y}}{4 Z_3 M_p^2} \left(1 \mp \sqrt{1
+ \frac{6Z T_3 V_2}{H_0^2 \sqrt{1+2Y}}} \right).
\end{eqnarray}
The inflation in this limit is parameterised by the slow roll term
\begin{equation}
\varepsilon \sim \exp \left(\frac{4 \delta N_e \sqrt{1+2Y}}{3(1+4Y)} \right),
\end{equation}
where we have reintroduced the parameter $\delta$ as in the $N=2$ section, which is once again given by
\begin{equation}
\delta = \frac{M_p^2 H_2}{H_0}. \nonumber
\end{equation}
The validity of the expression is determined by the background fluxes
manifest in the $Y$ terms.

Again we find a similar bound on $V_2$ as in the two-brane case, namely
\begin{equation}
V_2 \ge \frac{10^{-1}V_0}{M_p^2} F(Y)
\end{equation}
where the function $F(Y)$ now ranges between $F(Y) = 1 \to 2\sqrt{2Y}$.

In both cases we see that in order to satisfy the observed scalar curvature bound, we require
\begin{equation}
V_0 \le 1 \times 10^{-7} \left(\frac{M_p}{M_s} \right)^4
\end{equation}
which can again be used to constrain the maximal value of $V_2$. Once again we see that
the inflationary scale is sensitive to the magnitude of the string scale.
\section{Discussion}

In this paper we have examined the evolution of the IR DBI inflation model
in the context of the non-Abelian world-volume theory. The extra
stringy degrees of freedom present in the theory serve to induce a so-called
'fuzzy' potential term which makes the form of the action
significantly different from that of a single brane when we switch to
the finite $N$ theory.
We have seen that the large $N$ limit of the solution is remarkably similar
to the single brane case. This is not entirely unexpected since
this has a dual description in terms of a single classical object due to the Myers effect \cite{myers}. 
In fact we expect that our model to be the field theory dual of a supergravity solution consisting of a single $D5$-brane wrapping a non-trivial two-cycle with $N$ units of magnetic flux on the world volume \cite{blowup}. 
However in order for this configuration to be stable we will generally have to ensure that there is additional worldvolume electric flux
on the wrapped brane. From our field theory perspective, consistency of the two descriptions requires us to turn on a $U(1)$ electric
field on the coincident $D3$-branes, thus breaking the symemtry group $U(N) \to SU(N) \times U(1)$. 
It would be useful to investigate this theory and see if it does match up with our non-Abelian
one at large $N$, not only because of the intrinsic interest of how gauge fields affect cosmological behaviour, but because this may
also leads to new understanding of Yang-Mills theory via gauge /gravity duality.

Within the remit of our assumptions, we have tried to argue that inflation
is reasonably generic. By this we mean that we can generate at least
$60$ e-folds with relative ease due to large $N$ suppression.
However for the Ads type solutions we see that the scalar index cannot satisfy observational bounds and is thus immediately ruled out. The mass gap case is a little more subtle, because it can be a viable model provided we fine tune the solution.

The most intriguing results concern the analysis of the models where $N$ is finite.
Due to the highly nonlinear nature of these theories its is not clear what general conclusions we can draw. However we studied in detail the two examples of $N=2$ and $N=3$.
(It is certainly possible to analyse larger values of $N$, but the form of the action becomes more and more complicated). Already in these examples, we find dramatically different
behaviour from the cosmology of a single brane. Most notable is that the speed of sound
becomes a far more complicated function of the inflaton and we are restricted to certain
regions of the moduli space of solutions. 
For the $N=3$ case we noted that the speed of sound actually has zeros in its function. The physical explanation for this is not immediately obvious, but could lead to important physical predictions which distinguishes this model from all other string inspired inflationary scenarios. An obvious question is whether this phenomenon simply arises from the symmetrised trace conjecture, or whether it is an inherent feature of DBI models with an odd (not including 1)number of branes.

Another important result concerning this particular solution is that small values of $g_s KM$ are prefered which can be satisfied with a small string coupling constant and small fluxes.
If we can trust string perturbation theory then we can find a decent inflationary model. However for large fluxes this leads to large non-gaussianities, which may or may not be a good test of the model. Only observations from the Planck mission will resolve this issue.
Again one may ask whether this is an artifact of having an odd number of branes, as the $N=2$ case appears to have no such restriction.

We also found that the scalar spectral index can never be greater than unity in these models. In the case of mass gap backgrounds we found that eternal inflation appears to be a common feature of both $N=2$ and $N=3$ models.

Most of our results required us to make the simplifying assumption of $\gamma>>1$ in the 
non-Abelian DBI theory, in order to be able to carry out analytical computations. This assumption was also used in the Abelian DBI cosmology in \cite{tong}. 
However we were also able to consider the non-relativistic regime where $\gamma$ can be expanded in a power series about unity. In this limit we see
that inflation does indeed end in both types of background. However the speed of sound is essentially unity and therefore doesn't provide us with
a nice falsifiable prediction - since we can always tune the parameters to allow for enough inflation and suitably low scalar amplitude. Of course
this will inevitably change if we consider a full string compactification.
A more general analysis would probe the intermediate velocity spectrum, which we hope will be pursued in the future.

Our model differs significantly in some respects from a single probe brane,
despite sharing many similarities. The main
difference is that now the world-volume theory is playing the role of
the universe, rather than just the inflaton sector in the standard
IR approach to inflation. In fact this makes the model a hybrid between that
of DBI inflation \cite{tong} and Mirage Cosmology \cite{mirage}.
As a consequence one may ask about reheating in our model \footnote{Here is a partial list of papers which deal with 
reheating \cite{reheating}.}. We cannot
simply assume that there is another throat which contains the standard model
since this will not couple to the inflaton sector through open string modes.
One potential solution to this problem is to assume the existence of angled branes
in another throat. As our non-Abelian brane configuration travels up the throat
it will eventually feel a potential being sourced by these branes
and will move towards them. The standard model can be realised once
these branes intersect. Of course this implicitly assumes that we have
enough branes in our model to allow for a sufficiently large gauge
group to encompass the standard $SU(3) \times SU(2) \times U(1)$ gauge
symmetry of the standard model. An interesting point is that we could actually begin
with a model which has a large enough gauge group to contain the Standard Model, but still
within the realms of the finite $N$ expansion. In this case we may find that inflation direclty mixes with the observable sector, and a more 
comprehensive analysis using \cite{mazumdar} must be undertaken if we are to place the model on a firm experimental footing.
This is an interesting, but challenging subject which is left for future study in the context
of finite $N$ solutions. 
We remark that due to the nature of our approximations, we anticipate that there
is universal behaviour for any value of $N$ for some of the observables. For example we conjecture that the 
scalar index imposes the approximate WMAP bound
\begin{equation}
\frac{MK V_2}{g_s} \frac{M_s^4 10^{-7}}{(N+1)} \ge M_p^2
\end{equation}
which will clearly be harder to satisfy for larger values of $N$ without tuning. Once again note the presence of the
inflaton 'mass' term above.
This expression is an order of magnitude approximation, as we have anticipated that the five-dimensional
volume is $ \mathcal{O}(1) \pi^3$.

As regards future directions, it would also be useful to embed our model into a more realistic
compactification along the lines of \cite{kklt, fluxpotential},
by including the position of the D3-branes in the K\"ahler potential and
the non-perturbative superpotential. For the case of finite $N$,
one would hope that the closed string corrections would be small and computable
in a certain region of the moduli space (see \cite{haack} for the open string computation). 
The relevant scales will now be set by the specific choice of compactification.
Since the metric will now also explicitly contain couplings to the K\"ahler moduli, it seems likely that
DBI inflation will be more complicated in a fully compactified theory due to the fact that both the radion and the 
modulus will appear inside the DBI action. It may still be possible to obtain inflationary solutions in the single brane
case, although moduli stabilisation effects may reverse the effects or the warping. In any event, one would hope that
this could be tackled as a future problem.

It is also possible to extend the work initiated here to include world-volume
Abelian or non-Abelian gauge fields. It is an experimental
fact that we see weak magnetic fields at large scales in our universe today.
Although including a non zero B field complicates the
inflationary analysis, it will certainly play an important role in
a reheating phase. It would also be useful to consider an extension of \cite{giantinflaton} to the case of finite $N$.
Finally, another avenue of possibility would be to consider the coupling of
open string tachyons in the model, which have been shown to lead to
interesting inflationary solutions \cite{tachyon}.

\begin{center}
\textbf{Acknowledgements.}
\end{center}
We would like to thank Zong-Kuan Guo, Nobuyoshi Ohta and Shinji Tsujikawa for collaboration at an early stage
of this project, and also Paulo Moniz, Costis Papageorgakis, James Bedford and Massimo Giovannini for useful discussions.
JW is supported by a Queen Mary studentship and a Marie Curie Early Stage Training grant.
\section{Appendix}
In this section we explicitly calculate the relevant perturbation amplitudes
for the non-Abelian action. The definitions of the parameters in this section
differ from those in other sections in order to simplify the calculations
as much as possible - and we use units where $M_p=1$.
The action in the general case can is a non-linear function of 
the inflaton field and its time derivative, therefore it can be written in the following form
- consistent with the general prescription described in \cite{Hwang, mukhanov}
\begin{equation}
S=\int {\rm d}^4 x \sqrt{-g} \left[ \frac{R}{2}+p(\phi, X) \right]\,,
\end{equation}
where
\begin{eqnarray}
p=-NT_{3} \left[ h^4(\phi) \sqrt{1-2h^{-4}(\phi)T_3^{-1}X}
\sqrt{1+C^{-1}h^{-4}(\phi) \phi^4}-h^4(\phi)+V(\phi) \right]\,.
\end{eqnarray}
with $X=\dot{\phi}^2/2$ and $C=\lambda^2 \hat{C}T_3^2/4$.
We will explicitly consider the case of large $N$ in this appendix, however it is straightforward to show that the
derived results also apply for the finite $N$ case.
The background equations following from this are
\begin{eqnarray}
& & 3H^2=2Xp_{,X}-p \equiv \rho\,, \nonumber \\
& & \dot{H}=-Xp_{,X}\,, \nonumber \\
& & \frac{1}{a^3} (a^3 \dot{\phi} p_{,X})^{\cdot}
-p_{,\phi}=0\,.
\end{eqnarray}
Note that the energy density $\rho$ is given here by
\begin{eqnarray}
\rho=NT_{3} \left[ \frac{h^4(\phi) W(\phi)}
{\sqrt{1-2h^{-4}(\phi)T_3^{-1}X}}-h^4(\phi)
+V(\phi) \right]\,,
\end{eqnarray}
where the fuzzy potential is
\begin{equation}
W(\phi)=\sqrt{1+C^{-1}h^{-4}(\phi) \phi^4}\,.
\end{equation}

We consider the following general
perturbed metric about a FRW background
\begin{eqnarray}
\hspace*{-0.2em}\rd s^2 &=& - (1+2A)\rd t^2 +
2a\partial_iB \rd x^i\rd t
\nonumber\\
\hspace*{-0.2em}&& +a^2\left[ (1+2\psi)\delta_{ij}
+2\partial_{ij}E+2h_{ij}
\right] \rd x^i \rd x^j\,,
\end{eqnarray}
where $\partial_i$ represents the spatial partial derivative
$\partial/\partial x^i$ and
$\partial_{ij}=\nabla_i\nabla_j-(1/3)\delta_{ij}\nabla^2$.
Here $A$, $B$, $\psi$ and $E$ denote scalar metric
perturbations, whereas $h_{ij}$ represents tensor perturbations.
Defining the so-called comoving perturbation
\begin{eqnarray}
{\cal R} \equiv \psi-\frac{H}{\dot{\phi}}\delta \phi\,,
\end{eqnarray}
the Fourier modes of curvature perturbations
satisfy the following expression \cite{Hwang}
\begin{eqnarray}
\label{veq}
v''+\left( c_{S}^2 k^2 -\frac{z''}{z}
\right) v=0\,,
\end{eqnarray}
where
\begin{eqnarray}
\label{z2def}
& & z^2= \frac{a^2 \dot{\phi}^2 (p_{,X}+2Xp_{,XX})}
{H^2}\,,\nonumber \\
& & v=z {\cal R}\,,\nonumber \\
& & c_{S}^2=\frac{p_{,X}}{\rho_{,X}}
=\frac{p_{,X}}{p_{,X}+2Xp_{,XX}}\,.
\label{cRdef}
\end{eqnarray}
Note that $k$ is a comoving wavenumber and a prime represents
a derivative with respect to a conformal time $\tau=\int a^{-1} {\rm d}t$.
If the variable $z$ has a time-dependence  $z \propto |\tau|^q$,
one has $z''/z=\gamma_{S}/\tau^2$ with
$\gamma_{S}=q(q-1)$.
As long as $c_S^2$ is a positive constant or a slowly varying
positive function, the solution for (\ref{veq}) is given by
\begin{eqnarray}
v=\frac{\sqrt{\pi |\tau|}}{2}
\left[ c_1 (k) H_{\nu_{S}}^{(1)} (c_{S} k |\tau|)+
c_2 (k) H_{\nu_{S}}^{(2)} (c_{S} k |\tau|)
\right]\,,
\end{eqnarray}
where $\nu_{S}=\sqrt{\gamma_{S}+1/4}=|q-1/2|$.
The coefficients are chosen to be $c_1=0$ and $c_2=1$
to recover positive frequency solutions in a Minkowski
vacuum in an asymptotic past.

Defining the  spectrum of curvature perturbation as
${\cal P}_{\cal R}=k^3 |{\cal R}|^2/2\pi^2$, we obtain
\begin{eqnarray}
{\cal P}_{\cal R} &=& \frac{a^2 c_{S}^{-2\nu_{S}}}{z^2}
\left(\frac{H}{2\pi}\right)^2
\left(\frac{1}{aH|\tau|}\right)^2
\left(\frac{\Gamma(\nu_{S})}{\Gamma(3/2)}\right)^2
\left(\frac{k|\tau|}{2}\right)^{3-2\nu_{S}}
\nonumber \\
&\equiv&
\mathcal{A}_{S}^2 \left(\frac{k|\tau|}{2}\right)
^{3-2\nu_{S}},
\label{PS}
\end{eqnarray}
The spectral index of the power spectrum is
\begin{eqnarray}
n_{S}-1=3-2\nu_{S}=3-\sqrt{4\gamma_{S}+1}\,,
\end{eqnarray}
which means that the scale-invariant spectrum corresponds to
$\nu_{S}=3/2$.
About the de-Sitter background with $|\tau|=1/aH$,
the amplitude of the curvature perturbation is given by
\begin{eqnarray}\label{eq:curvature}
\mathcal{A}_{S}^2 \simeq \frac{1}{p_{,X}c_{S}}
\left( \frac{H^2}{2\pi \dot{\phi}} \right)^2\,.
\end{eqnarray}
The tensor perturbations satisfy the same equation as in the case
of standard slow-roll inflation.
Taking into account polarization states of tensor modes,
The power spectrum is given by
\begin{eqnarray}
{\cal P}_T &=&
8 \left(\frac{H}{2\pi}\right)^2
\left(\frac{1}{aH|\tau|}\right)^2
\left(\frac{\Gamma(\nu_{T})}{\Gamma(3/2)}\right)^2
\left(\frac{k|\tau|}{2}\right)^{3-2\nu_T}
\nonumber \\
&\equiv&
\mathcal{A}_T^2 \left(\frac{k|\tau|}{2}\right)
^{3-2\nu_T},
\label{PT}
\end{eqnarray}
where $\nu_{T}=\sqrt{\gamma_T+1/4}$ with
$a''/a=\gamma_{T}/\tau^2$.
Hence about the de-Sitter background the amplitude of
the tensor perturbation is
\begin{eqnarray}\label{eq:tensor}
\mathcal{A}_T^2 \simeq 8 \left(\frac{H}{2\pi}\right)^2\,.
\end{eqnarray}
The spectral index of the power spectrum is
\begin{eqnarray}
n_T=3-2\nu_T=3-\sqrt{4\gamma_T+1}\,.
\end{eqnarray}
The tensor to scalar ratio is
\begin{eqnarray}\label{eq:ratio}
r=\frac{\mathcal{A}_T^2}{\mathcal{A}_S^2}=8\frac{\dot{\phi}^2}{H^2}
p_{,X} c_{S}\,.
\label{ratio}
\end{eqnarray}
To study the running of the spectral indices we find it convenient to introduce
the following parameters:
\begin{eqnarray}
\epsilon_{1}=-\frac{\dot{H}}{H^2}\,,\quad
\epsilon_{2}=\frac{\ddot{\phi}}{H \dot{\phi}}\,,\quad
\epsilon_{3}=\frac{\dot{F}}{2HF}\,.
\end{eqnarray}
where $F \equiv p_{,X}+2Xp_{,XX}$ and $\epsilon_1$ is the same as
the $\varepsilon$ which we use in the main part of the paper. If
$\dot{\epsilon}_{i}=0$, we can derive
\begin{eqnarray}
\frac{z''}{z}=\frac{\gamma_{\cal R}}{\tau^2}\,,\quad
\gamma_{\cal R}=
\frac{(1+\epsilon_1+\epsilon_2+\epsilon_3)(2+\epsilon_2+\epsilon_3)}
{(1-\epsilon_1)^2}\,.
\end{eqnarray}
Under the slow-roll approximation $|\epsilon_i| \ll 1$, we find that
the spectral index of the curvature perturbation is given by
\begin{eqnarray}
n_{S}-1=-2(2\epsilon_1+\epsilon_2+\epsilon_3)\,.
\end{eqnarray}
Similarly the spectral index of the tensor perturbation is
\begin{eqnarray}
n_T=-2\epsilon_1\,.
\end{eqnarray}
By using the background equations we have
$\epsilon_1=\dot{\phi}^2 p_{,X}/(2H^2)$.
This then shows that the tensor to scalar ratio (\ref{eq:ratio})
yields
\begin{eqnarray}
r=16 \epsilon_1 c_{S}=-8c_S n_{T}\,.
\end{eqnarray}
Again this is the same expression as in the single brane case, and is
a distinctive feature of DBI inflation.
The WMAP normalisation we will employ in this paper
are the following \cite{wmap, braneconstraints}
\begin{eqnarray}
\mathcal{A_S}^2 &=& 10^{-9} \nonumber \\
r & = & \frac{\mathcal{A}_T^2}{\mathcal{A}_S^2} \le 0.55 \hspace{0.4cm} (\le 0.24 \hspace{0.2cm} \rm{at} \hspace{0.2cm} 0.95 \hspace{0.2cm} \rm{C.L}) \nonumber\\
n_s &=& 0.987^{+0.019}_{-0.037} \nonumber \\
f_{NL} &\le& 100 
\end{eqnarray}
which may differ slightly from normalisation used elsewhere.


\end{document}